\documentclass[conference]{IEEEtran}
\IEEEoverridecommandlockouts

\usepackage{cite}
\usepackage{amsmath,amssymb,amsfonts}
\usepackage{algorithmic}
\usepackage{graphicx}
\usepackage{textcomp}
\usepackage{xcolor}
\usepackage{booktabs}
\usepackage{tabularx}
\usepackage[utf8]{inputenc}
\usepackage{newunicodechar}
\newunicodechar{，}{,}
\usepackage{braket}

\usepackage{graphicx}
\usepackage{subcaption}
\usepackage[linesnumbered,algoruled,boxed,lined]{algorithm2e} 
\def\BibTeX{{\rm B\kern-.05em{\sc i\kern-.025em b}\kern-.08em
    T\kern-.1667em\lower.7ex\hbox{E}\kern-.125emX}}
\begin{document}

\title{Resource-Aware Grover Search for Minimum Vertex Cover\\
}

\author{
    \IEEEauthorblockN{Beilei Jiang, Harry Fu, Pavan Krishna Yarlagadda, Alexander Shan, Yunhe Feng, and Song Fu}
    \IEEEauthorblockA{\textit{University of North Texas, Denton, TX, USA} \\
    Email: \{beilei.jiang, yunhe.feng, song.fu\}@unt.edu, \\
    \{harryfu, pavankrishnayarlagadda, alexandershan\}@my.unt.edu}
}

\maketitle

\begin{abstract}
The Minimum Vertex Cover (MVC) problem is a fundamental NP-hard combinatorial optimization problem with applications in network analysis and resource allocation. Grover’s algorithm provides a quadratic reduction in the number of queries required for unstructured search, but existing Grover-based MVC formulations can incur substantial quantum resource overhead due to costly vertex-counting circuits and complex oracle constructions.

In this work, we develop and evaluate several encoding and oracle-design strategies for reducing the qubit count, circuit depth, and gate complexity of Grover-based MVC search. We first construct a Dicke-Parallel formulation that restricts the search to fixed-cardinality subsets, eliminating explicit vertex counting, together with a parallel edge-verification oracle that reduces feasibility-checking overhead. Building on this formulation, we develop an edge-counting oracle that replaces per-edge auxiliary storage with a logarithmic-size counting register, substantially reducing ancillary-qubit requirements. We further propose an edge-centric encoding that represents endpoint selections directly and derives vertex-selection states through incident-edge Boolean operations, enabling more depth-efficient oracle construction.

Our resource analysis reveals complementary trade-offs among qubit width, circuit depth, gate count and Grover iteration count. The Edge-Counting formulation is particularly attractive under tight qubit constraints, whereas the
Edge-Centric formulation can reduce both circuit depth and Grover iteration count for graphs whose overall edge count remains moderate while their minimum vertex covers contain sufficiently many internal edges. Such structures increase the number of edge assignments representing each valid cover and can therefore increase the marked-state fraction of the Edge-Centric search. Sparse or moderately sparse graphs with suitable local structures, such as pendant or low-degree regions, can exhibit this advantage. In contrast, the Dicke-Parallel formulation provides a more robust choice when such representation multiplicity is limited or when graph density makes the edge-based search space large. These results provide practical guidance for selecting Grover-based MVC formulations according to hardware constraints and graph structure.

\end{abstract}


\section{Introduction}

The Minimum Vertex Cover (MVC) problem is a fundamental combinatorial optimization problem with wide-ranging applications in network analysis, scheduling, and resource allocation~\cite{network,network1, scheduling,resource,resource2, resource3}.
Given an undirected graph $G=(V,E)$, the objective is to find the smallest subset of vertices $V' \subseteq V$ such that every edge in $E$ has at least one endpoint in $V'$.
Because MVC is NP-hard, exact solution methods become computationally prohibitive for large graphs, motivating the exploration of alternative computational paradigms.

Quantum computing provides new opportunities for accelerating combinatorial search.
Among quantum algorithms, Grover's algorithm~\cite{grover1,grover2,grover4,grover5,grover6} provides a quadratic speedup over classical exhaustive search for unstructured search problems, making it a natural candidate for NP-hard optimization problems such as MVC.
Prior Grover-based MVC formulations encode candidate vertex subsets into quantum states and construct an oracle that verifies both edge coverage and solution cardinality~\cite{baseline,10383108}.
However, enforcing these constraints can require substantial quantum resources.
In particular, dynamic-programming-based vertex counting introduces significant auxiliary-qubit overhead and can require $\mathcal{O}(n^2)$ qubits for an $n$-vertex graph~\cite{baseline}.

Dicke-state-based formulations provide an alternative way to enforce the cardinality constraint.
For the $k$-vertex cover problem (k-VCP), a Dicke state restricts the search space to computational basis states with exactly $k$ selected vertices~\cite{dicke_vc}, thereby eliminating the need for explicit in-circuit vertex counting.
However, removing cardinality counting does not eliminate the cost of feasibility verification.
Existing Dicke-state-based designs may still require substantial auxiliary state, such as replicated vertex information or per-edge flag qubits, and repeated multi-controlled operations can lead to considerable circuit depth.
Therefore, while Dicke-state restriction addresses the cardinality-counting overhead, efficient edge-coverage verification remains an important challenge.

More broadly, Grover-based MVC implementations contain multiple sources of quantum resource overhead.
Cardinality evaluation can dominate the qubit footprint in explicit counting formulations, while feasibility verification can introduce additional ancilla and circuit depth as the number of edges increases.
These costs need not be addressed by a single oracle organization.
Instead, different architectural choices can target different resource constraints, ranging from balanced designs that reduce both counting and verification overhead to more specialized designs that prioritize either qubit footprint or circuit depth.

In this work, we develop three complementary Grover-based MVC designs targeting different quantum-resource regimes: Dicke-Parallel, Edge-Counting, and Edge-Centric. Dicke-Parallel combines fixed-weight Dicke-state search with a parallel edge-verification oracle, eliminating explicit vertex counting while reducing feasibility-checking overhead. Edge-Counting replaces per-edge auxiliary storage with a shared reversible counter, substantially reducing auxiliary-qubit requirements at the cost of additional sequential computation. Edge-Centric reformulates the search using edge-based endpoint selections and derives vertex states through incident-edge Boolean operations, enabling highly parallel cardinality evaluation and reduced circuit depth.







These three designs occupy different regions of the quantum-resource design space. Dicke-Parallel provides a balanced and robust formulation, Edge-Counting is particularly attractive under tight qubit constraints, and Edge-Centric can provide substantial depth and iteration-count advantages for suitable sparse or moderately sparse graph structures.

The rest of the paper is organized as follows.
Section~\ref{background} describes the Minimum Vertex Cover problem， Grover's algorithm and reviews related work.
Section~\ref{motivation} analyzes the resource limitations of existing Grover-based MVC algorithms.
Sections~\ref{design} and~\ref{evaluation} present and evaluate the proposed designs.
Finally, Section~\ref{conclusion} concludes the paper.

\section{Background and Related Work}
\label{background}

\subsection{Minimum Vertex Cover}

Given an undirected graph $G=(V,E)$, a vertex cover is a subset of vertices $V' \subseteq V$ such that every edge in $E$ has at least one endpoint in $V'$. Specifically, for each edge $(u,v)\in E$, either $u\in V'$ or $v\in V'$ must hold. A minimum vertex cover (MVC) is a vertex cover with the smallest cardinality among all valid vertex covers of $G$.

As a simple example, consider a graph $G=(V,E)$ with vertex set
$V=\{0,1,2,3\}$ and edge set $E=\{(0,1),(1,2),(1,3),(2,3)\}$. For this graph, both $\{1,2\}$ and $\{1,3\}$ are minimum vertex covers of size two.

MVC is NP-hard, and its decision version is NP-complete. Although a candidate vertex cover can be verified in polynomial time, no polynomial-time algorithm is known for finding an optimal solution on general graphs unless ${P}={NP}$. As a result, practical approaches to MVC generally rely on exact exponential-time algorithms, approximation algorithms, or heuristics~\cite{mvc1,yamout2022parallel,mvc2,mvc3,mvc4,heuristic}.

Exact approaches, such as branch-and-bound and bounded-search-tree algorithms, can find optimal solutions but exhibit exponential worst-case complexity~\cite{mvc1,yamout2022parallel}. Approximation algorithms can scale to larger instances while sacrificing optimality. For example, classical algorithms based on maximal matching or linear-programming relaxation can achieve a factor-2 approximation in polynomial time~\cite{mvc2}. Heuristic and local-search approaches often perform well on large practical instances but do not provide worst-case guarantees of optimality~\cite{mvc3,mvc4,heuristic}.

The combinatorial search complexity of MVC makes it a natural target for quantum search-based approaches, which may reduce the number of oracle evaluations required to identify a valid or optimal solution.

\subsection{Grover's Algorithm}

Grover's algorithm~\cite{grover1,grover2,grover3} provides a quadratic speedup for unstructured search by amplifying the probability amplitudes of marked states while suppressing those of non-solution states. Starting from an initial superposition over the search space, the algorithm repeatedly applies Grover iterations that rotate the quantum state toward the subspace containing valid solutions.

Each Grover iteration consists of two main components: a
\emph{phase oracle} and a \emph{diffusion operator}. The phase oracle marks solution states by applying a selective phase inversion, while the diffusion operator performs a reflection that amplifies the amplitudes of the marked states.

For a search space of size $N$ containing $M$ valid solutions, Grover's algorithm requires approximately $\mathcal{O}(\sqrt{N/M})$ oracle applications, compared with an expected $\mathcal{O}(N/M)$ evaluations for classical random search. However, this query-complexity advantage does not directly determine the practical cost of a quantum implementation. For combinatorial problems such as MVC, a substantial portion of the quantum resources can be consumed by constructing the oracle that determines whether a candidate state satisfies the problem constraints.

In the MVC problem, such an oracle generally needs to evaluate two conditions. First, the selected vertices must satisfy the \emph{feasibility constraint}: every edge must have at least one selected endpoint. Second, the candidate must satisfy a \emph{cardinality constraint}, corresponding to the desired number of selected vertices. The implementation of these two operations is therefore central to the resource efficiency of Grover-based MVC algorithms.

\subsection{Grover-Based Quantum Algorithms for MVC}

Previous Grover-based MVC formulations typically encode candidate vertex subsets directly in an $n$-qubit vertex register and construct an oracle that jointly evaluates edge coverage and vertex cardinality~\cite{baseline,10383108}. Although Grover search reduces the number of oracle evaluations required to find a valid solution, implementing these constraints reversibly can introduce substantial auxiliary-qubit and circuit overhead.

Wang et al.~\cite{baseline} proposed a Grover-based framework in which the cardinality constraint is enforced through a
dynamic-programming-style vertex-counting circuit. The circuit determines whether the number of selected vertices equals the desired cover size while the remaining oracle logic verifies edge coverage. This formulation provides a direct implementation of the MVC constraints within the Grover oracle, but the explicit counting structure introduces substantial auxiliary state. In particular, the dynamic programming formulation can require $\mathcal{O}(n^2)$ qubits as the number of vertices grows.

Consequently, the resource cost of a Grover-based MVC algorithm is not determined only by the size of its search space. The internal organization of the oracle, particularly the mechanism used to enforce the cardinality constraint, can become a major scalability bottleneck. This motivates approaches that avoid explicit dynamic-programming-based vertex counting.

\subsection{Dicke-State-Based Quantum Search}

A different approach is to restrict the quantum search space to
candidate subsets that already satisfy the required cardinality.
Dicke states provide a natural representation for this purpose. 
An $n$-qubit Dicke state of Hamming weight $k$ is a superposition of computational basis states containing exactly $k$ qubits in the $\ket{1}$ state. Deterministic circuits for preparing Dicke states were introduced by B\"artschi and Eidenbenz~\cite{dicke}, enabling their use in quantum algorithms with fixed-cardinality constraints.

For the $k$-vertex cover problem (k-VCP), Dicke-state preparation can therefore eliminate the need to count selected vertices inside the oracle. Rather than searching over all $2^n$ vertex subsets, the algorithm searches only among the
$\binom{n}{k}$ subsets containing exactly $k$ selected vertices.

Building on this idea, Jiang et al.~\cite{dicke_vc} proposed the DSQS method, which prepares a fixed-weight superposition and applies an oracle to identify candidate states satisfying the vertex-cover constraint. By encoding the cardinality requirement directly into the initial state, this approach avoids the dynamic-programming-style vertex-counting circuitry required by earlier Grover-based formulations.

However, eliminating explicit cardinality counting does not eliminate the resource cost of feasibility verification. The Dicke-state-based design still requires auxiliary quantum state to evaluate whether all edges are covered. In particular, its edge-verification procedure uses per-edge flag structures together with a mirror vertex register for solution readout, followed by multi-controlled operations to aggregate the edge-coverage results. As the number of edges increases, these structures can contribute substantial qubit, gate, and circuit-depth overhead.

Importantly, this does not imply that Dicke-state-based formulations necessarily have greater circuit depth than
dynamic-programming-based formulations. Rather, the two approaches remove or expose different sources of resource cost. Explicit counting formulations incur substantial overhead in cardinality evaluation, whereas Dicke-state formulations eliminate this counting requirement but leave the cost of edge-feasibility verification as an important optimization opportunity.





\section{Resource Bottlenecks in Grover-Based MVC} \label{motivation}
Existing Grover-based MVC formulations expose three major resource bottlenecks: search-space size, oracle width, and oracle depth. These bottlenecks arise differently in conventional vertex-based and Dicke-state-based formulations and motivate the complementary designs developed in this work.

\subsection{Search-Space Size}
A conventional Grover-based MVC formulation represents a candidate solution using one qubit per vertex. The search therefore spans all possible vertex subsets,

\[
N_{{Base}} = 2^n.
\]

For a fixed candidate cover size ($k$), however, only subsets containing exactly $k$ vertices need to be considered. A Dicke-state formulation restricts the initial state to this fixed-cardinality subspace,

\[
N_{{Dicke}} = \binom{n}{k},
\]

thereby excluding all candidate states whose Hamming weight differs from $k$ before Grover search begins.

This reduction affects not only the size of the represented search space but also the number of Grover iterations. If $M_k$ minimum vertex covers of size $k$ exist, the corresponding iteration counts scale approximately as

\[
R_{{Base}}
=
\mathcal{O}
\left(
\sqrt{\frac{2^n}{M_k}}
\right),
\]

and

\[
R_{{Dicke}}
=
\mathcal{O}
\left(
\sqrt{\frac{\binom{n}{k}}{M_k}}
\right).
\]

Since

\[
\binom{n}{k} \leq 2^n,
\]

fixed-cardinality search provides a systematic reduction in the Grover search space. However, reducing the search space alone does not determine the total implementation cost, because the oracle used to identify feasible covers may itself require substantial quantum resources.

\subsection{Oracle Width}
In conventional Grover-based MVC formulations, the oracle must evaluate both edge feasibility and vertex cardinality. A major source of width overhead is the explicit cardinality-counting circuit.

For example, dynamic-programming-style counting formulations introduce auxiliary states $w_i^j$ to represent whether exactly $j$ of the first $i$ vertices have been selected. These states form a triangular dependency structure. For a graph with three vertices, the recurrence takes the form

\[
w_1^0 = o \wedge \overline{v_0},
\qquad
w_1^1 = o \wedge v_0,
\]

followed by

\[
\begin{aligned}
w_2^{0} &= w_1^{0} \wedge \overline{v_1}, \\
w_2^{1} &= (w_1^{0} \wedge v_1)
        \vee (w_1^{1} \wedge \overline{v_1}), \\
w_2^{2} &= w_1^{1} \wedge v_1,
\end{aligned}
\]

with the same construction repeated for subsequent vertices.

The resulting counting structure contains

\[
\mathcal{O}(n^2)
\]

intermediate states. In the formulation of~\cite{baseline}, the total qubit requirement is

\[
n + m + \frac{(n+1)(n+2)}{2} + 1,
\]
where $n$ and $m$ denote the number of vertices and edges, respectively. Thus, the explicit cardinality representation can become the dominant source of qubit overhead.

Dicke-state search removes this counting structure by enforcing the desired cardinality in the initial state. However, prior Dicke-state-based formulations still require auxiliary state for edge-feasibility verification, including per-edge flags and additional vertex-related registers~\cite{dicke_vc}. Therefore, eliminating explicit vertex counting reduces one source of width overhead but does not remove the need for resource-efficient feasibility checking.

This motivates oracle designs that explicitly trade auxiliary width against additional computation.

\subsection{Oracle / Grover-Iteration Depth} \label{Tdepth}
Circuit depth introduces a second, distinct bottleneck. In dynamic-programming-style counting circuits, later counting states depend recursively on earlier ones, creating sequential dependencies that limit parallel execution. The counting procedure can therefore incur superlinear depth as the graph size increases.

Dicke-state formulations eliminate this recursive cardinality computation, but the feasibility oracle can still contain repeated multi-controlled operations and sequential edge checks. As the number of edges grows, these operations can dominate the depth of a Grover iteration.

The relevant search cost is therefore not determined by the per-iteration circuit depth alone. If $D_{{iter}}$ denotes the depth of one Grover iteration and $R$ denotes the required number of iterations, the overall search depth is approximately

\[
D_{search}
\approx
R D_{iter}.
\]

Consequently, a formulation may reduce the depth of a single oracle while still requiring more Grover iterations, or vice versa. The search representation and the oracle structure must therefore be considered jointly.

These observations motivate the three complementary designs explored in this work. \emph{Dicke-Parallel} combines fixed-weight Dicke-state search with parallel edge verification to provide a balanced implementation. \emph{Edge-Counting} replaces per-edge auxiliary storage with a shared reversible counting mechanism, reducing qubit width at the cost of additional sequential computation. \emph{Edge-Centric} changes the search representation itself to expose greater parallelism in cardinality evaluation, while its edge-based search space introduces a graph-dependent trade-off in the number of Grover iterations. Together, these designs target different resource bottlenecks and motivate the comparative analysis presented in the following sections.

\section{Resource-Aware Grover Designs for MVC} \label{design}

\begin{algorithm}[t]
\caption{Dicke-Parallel Grover Search for MVC}
\label{alg:dicke_parallel}
\begin{algorithmic}[1]

\STATE \textbf{Input:} Graph $G=(V,E)$ with $|V|=n$, target cover size $k$, Grover iteration count $R_k$
\STATE \textbf{Output:} A vertex cover $C$ of size $k$

\STATE Allocate an $n$-qubit vertex register $Q_V$
\STATE Allocate one auxiliary qubit $q_j$ for each edge $e_j\in E$
\STATE Allocate a phase ancilla $q_p$ and initialize it to $\ket{-}$

\STATE Prepare the Dicke state $D_n^k$ on $Q_V$

\FOR{$r=1$ to $R_k$}

    \STATE \textit{// Edge-coverage evaluation:}
    \FORALL{$e_j=(u,v)\in E$}
        \STATE Compute $q_j \leftarrow (u \lor v)$
    \ENDFOR

    \STATE \textit{// Phase marking:}
    \STATE Apply a multi-controlled phase operation controlled by all
    edge ancillas $\{q_j\}$

    \STATE \textit{// Uncomputation:}
    \FORALL{$e_j=(u,v)\in E$}
        \STATE Uncompute $q_j \leftarrow (u \lor v)$
    \ENDFOR

    \STATE \textit{// Dicke-space diffusion:}
    \STATE Apply $(D_n^k)^\dagger$
    \STATE Apply the reflection about the reference state
    \STATE Apply $D_n^k$

\ENDFOR

\STATE Measure $Q_V$ to obtain candidate set $C$
\RETURN $C$


\end{algorithmic}
\end{algorithm}

\begin{figure*}[t]
    \centerline{\includegraphics[width=0.9\textwidth]{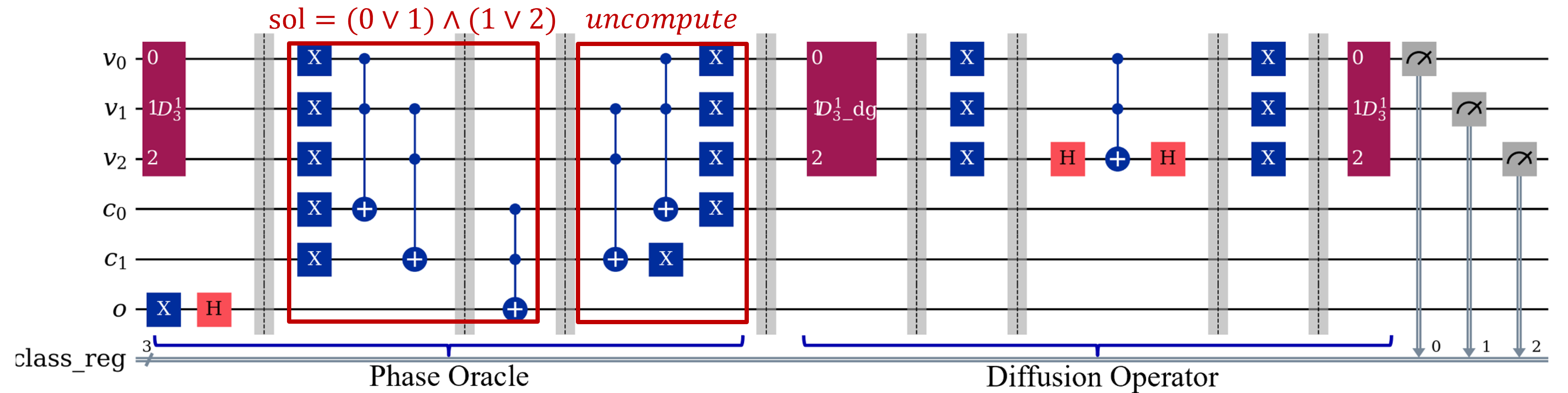}}
    \caption{Quantum circuit of the proposed Dicke-Parallel Grover search for MVC.}
    \label{fig:dicke_circuit}
    \vspace{-0.15 in}
\end{figure*}

In this section, we present three Grover-based formulations for the Minimum Vertex Cover (MVC) problem—Dicke-Parallel, Edge-Counting, and Edge-Centric—and examine their trade-offs in qubit usage, circuit depth, gate count and Grover iteration count.

\begin{enumerate}
    \item \textbf{Dicke-Parallel}, which eliminates expensive vertex-counting circuits while reducing oracle complexity.
    \item \textbf{Edge-Counting}, which reduces auxiliary qubit usage through logarithmic-size counting registers.
    \item \textbf{Edge-Centric}, which directly encodes edge coverage to achieve low-depth circuit design.
\end{enumerate}

\subsection{Dicke-Parallel}

As discussed in Section~\ref{motivation}, existing Grover-based MVC approaches typically reduce vertex counting complexity by adopting Dicke-state-based formulations, which eliminate dynamic-programming-style counting.
However, prior Dicke-state-based implementations instantiate this framework using relatively complex oracle constructions. In these designs, edge coverage verification is performed using additional flag qubits, while vertex mirror qubits are introduced to represent vertex selection states under the constrained subspace, together with multi-controlled operations applied in a sequential manner to enforce the required logical constraints within the oracle construction.

To address this, we construct an Dicke-Parallel formulation that combines Dicke-state-based search space restriction with a parallelized edge-coverage verification oracle, aiming to reduce both qubit overhead and circuit complexity within a unified design.

Algorithm~\ref{alg:dicke_parallel} summarizes the Dicke-Parallel formulation for a fixed target cardinality \(k\). The Dicke state restricts the candidate space to weight-\(k\) vertex subsets, while the oracle evaluates all edge-coverage predicates using separate edge ancillas before jointly applying the phase mark.

Specifically, the Dicke state is employed to eliminate the need for explicit vertex counting. Candidate solutions are explored in order of increasing cardinality by preparing Dicke states $D_n^k$, where $n$ is the number of vertices and $k$ is the number of selected vertices. For example, for three vertices, the Dicke state with $k=1$ is:
\[
D_3^1 = \frac{1}{\sqrt{3}} (|001\rangle + |010\rangle + |100\rangle).
\]

Unlike prior Dicke-state-based approaches that use a two-flag-per-edge construction and a mirror vertex register for solution readout, our oracle evaluates each edge-coverage predicate directly using a single auxiliary qubit per edge. Since these edge predicates are independent local constraints, they can be evaluated in parallel.  
As illustrated in Figure~\ref{fig:dicke_circuit}, for a graph with edges $(v_0,v_1)$ and $(v_1,v_2)$, the vertex cover condition is expressed as
\[
(0 \vee 1) \wedge (1 \vee 2),
\]
where OR operations are implemented using anti-Toffoli gates and AND operations using Toffoli gates.

The oracle output qubit $o$ is set to $|1\rangle$ if and only if the selected vertex subset covers all edges. For phase marking, $o$ is initialized to $|-\rangle$, enabling phase inversion of valid solutions.

The diffusion operator is then applied to amplify the probability amplitudes of the marked states, followed by measurement of the vertex qubits.

\subsection{Edge-Counting}
\begin{algorithm}[t]
\caption{Edge-Counting Grover Search for MVC}
\label{alg:edge_counting}
\begin{algorithmic}[1]

\STATE \textbf{Input:} Graph $G=(V,E)$ with $|V|=n$ and $|E|=m$,
target cover size $k$, Grover iteration count $R_k$
\STATE \textbf{Output:} A vertex cover $C$ of size $k$, if one exists

\STATE Allocate an $n$-qubit vertex register $Q_V$
\STATE Allocate a temporary edge-coverage qubit $q_e$
\STATE Allocate a counter register $Q_W$ of
$\lceil \log_2(m+1) \rceil$ qubits
\STATE Allocate a phase ancilla $q_p$ and initialize it to $\ket{-}$

\STATE Prepare the Dicke state $D_n^k$ on $Q_V$

\FOR{$r=1$ to $R_k$}

    \STATE \textit{// Edge counting}
    \FOR{each edge $e_j=(u,v)\in E$}
        \STATE Compute $q_e \leftarrow (u \lor v)$
        \STATE Conditionally increment $Q_W$ by $q_e$
        \STATE Uncompute $q_e$
    \ENDFOR

    \STATE \textit{// Phase marking}
    \IF{$Q_W=m$}
        \STATE Apply a phase flip using $q_p$
    \ENDIF

    \STATE \textit{// Uncomputation}
    \STATE Reverse the edge-counting procedure to restore
    $Q_W$ and $q_e$ to $\ket{0}$

    \STATE \textit{// Dicke-space diffusion}
    \STATE Apply $(D_n^k)^\dagger$
    \STATE Apply the reflection about the reference state
    \STATE Apply $D_n^k$

\ENDFOR

\STATE Measure $Q_V$ to obtain candidate set $C$
\RETURN $C$

\end{algorithmic}
\end{algorithm}

\begin{figure*}[t]
    \centerline{\includegraphics[width=1\textwidth]{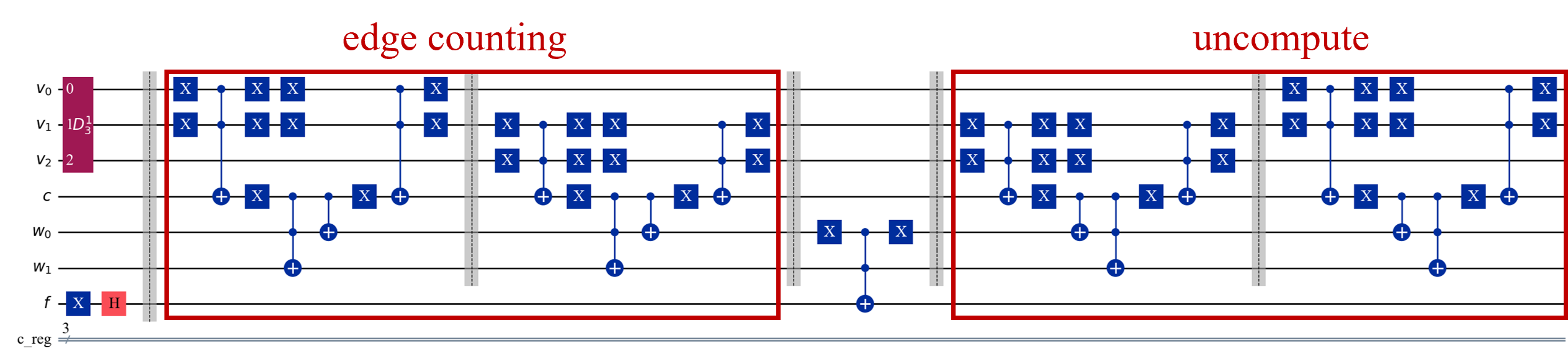}}
    \caption{Quantum circuit of the proposed Edge-Counting Grover search for MVC.}
    \label{fig:counting}
\end{figure*}

\begin{figure}
    \centerline{\includegraphics[width=0.42\textwidth]{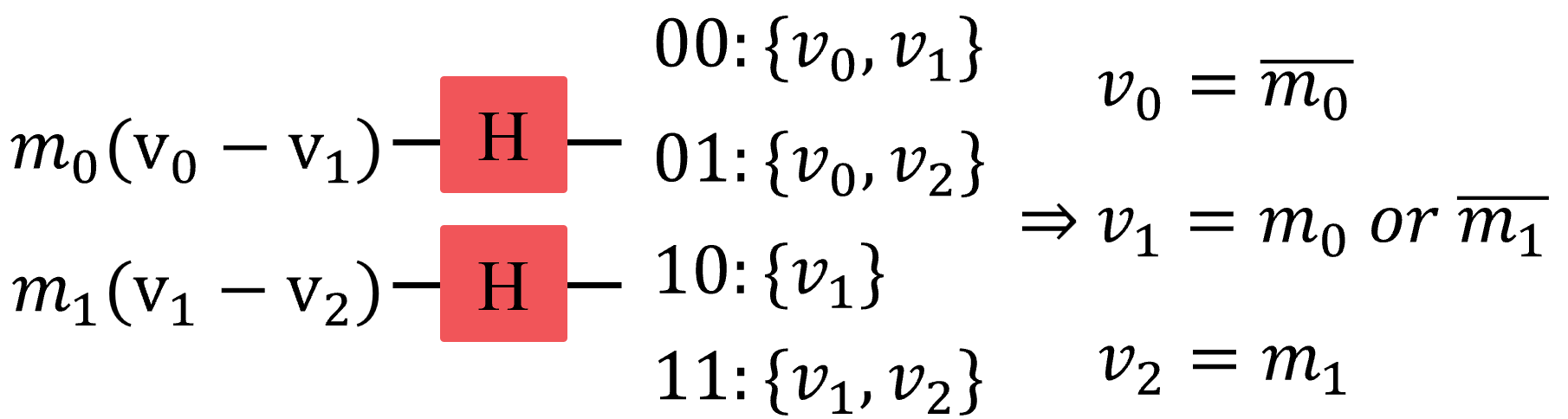}}
    \caption{Illustration of edge-centric encoding and vertex-state derivation.}
    \label{fig:edgeEn}
\end{figure}

To further reduce qubit usage, we propose a \textbf{counting-based strategy} in the phase oracle: instead of using auxiliary qubits to record each edge's coverage, we accumulate the total number of covered edges for a vertex subset.

A vertex subset corresponds to a valid MVC solution if and only if it covers all $m$ edges. Therefore, the oracle only needs to verify whether the total number of covered edges equals $m$, rather than checking each edge independently.

Algorithm~\ref{alg:edge_counting} summarizes the Edge-Counting formulation for a fixed target cardinality \(k\). The Dicke state restricts the candidate space to weight-\(k\) vertex subsets, while the edge-counting oracle calculte the number of edges covered before applying the phase mark.

Specifically, the oracle operates by sequentially processing each edge $(u,v)$. For every edge, a reversible logic block first determines whether at least one of its endpoints is selected. If the edge is covered, a controlled increment operation is applied to a binary counter encoded in qubits $c_{\ell-1}\cdots c_1 c_0$, where $\ell = \lfloor \log_2 m \rfloor + 1$.

The counter is implemented using a ripple-carry adder structure, where each increment propagates carries from the most significant bit to lower-order bits as needed. 

As illustrated in Figure~\ref{fig:counting}, for a graph with edges $(0,1)$ and $(1,2)$, when evaluating edge $(0,1)$, if vertex $0$, vertex $1$, or both are selected, the counter is incremented from $00$ to $01$, indicating that one edge is covered. When processing the next edge $(1,2)$, if it is also covered, the counter is incremented to $10$, representing two covered edges. A comparator is employed then to check whether the counter value equals $m$, 2 in this example. If the equality condition is satisfied, the oracle flips the output qubit, implementing a $-1$ phase factor. Finally, all intermediate computations, including the counter and ancilla qubits, are uncomputed to restore the ancilla registers to their initial states.

For clarity, the diffusion operator is omitted from Figure~\ref{fig:counting}. It is applied after phase marking to amplify the amplitudes of the marked states, followed by measurement of the vertex qubits.



\subsection{Edge-Centric}


Although Edge-Counting substantially reduces qubit usage, the sequential accumulation of edge-coverage information introduces additional gate overhead and circuit depth. To reduce this depth bottleneck, we explore an alternative Edge-Centric encoding that represents endpoint selections at the edge level and derives the corresponding vertex-selection states from these edge variables.

\subsubsection{Encoding definition}
Let each edge $e_i=(u_i, v_i) \in E$ be associated with one binary edge variable $m_i$. The value of $m_i$ specifies which endpoint of $e_i$ is selected:
\[
m_i = 0 \Rightarrow u_i, \quad m_i = 1 \Rightarrow v_i
\]
Therefore, an $m$-edge graph is represented by an $m$-qubit search register, whose computational basis states correspond to endpoint-selection assignments over all edges. Because every edge explicitly selects one of its two endpoints, each basis state induces a valid vertex cover by construction.

\subsubsection{Boolean Derivation of Vertex States} \label{vetex_derivation}
The vertex-selection state of each vertex is derived from the edge variables associated with its incident edges. For an edge $e_i=(u_i,v_i)$, the variable $m_i$ contributes $\overline{m_i}$ to endpoint $u_i$ and $m_i$ to endpoint $v_i$. The state of a vertex is obtained by OR-ing all such literals contributed by its incident edges.

As illustrated in Figure~\ref{fig:edgeEn}, consider a graph $G=(V,E)$ with 
\[
V=\{0,1,2\}, \quad E=\{(0,1),(1,2)\}. 
\]
Two qubits $m_0$ and $m_1$ encode edges $(v_0,v_1)$ and $(v_1,v_2)$, respectively. Applying Hadamard gates to the edge register produces an equal superposition over all possible edge-status configurations. For each configuration, the corresponding vertex-selection state is then derived according to the encoding rule defined above. 
The four basis states induce the following vertex subsets:
\[
\begin{aligned}[t]
|00\rangle &\rightarrow \{v_0,v_1\}, \\
|01\rangle &\rightarrow \{v_0,v_2\}, \\
|10\rangle &\rightarrow \{v_1\}, \\
|11\rangle &\rightarrow \{v_1,v_2\}.
\end{aligned}
\]

For this example, the corresponding vertex-selection functions are
\[
v_0 = \overline{m_0}, \quad v_1 = m_0 \vee \overline{m_1}, \quad v_2 = m_1.
\]
Thus, the edge register directly determines the corresponding vertex-selection state without requiring a separate edge-feasibility test.

\subsubsection{Validity and Reachability}
The edge-centric representation must satisfy two properties to support MVC search. First, every edge assignment should induce a valid vertex cover. Second, every minimum vertex cover should be representable by at least one edge assignment. We establish these properties below.

\textbf{Proposition I.}
Every computational-basis state in the edge-centric representation induces a valid vertex cover.

\textit{Proof.}
For each edge $e_i=(u_i,v_i)$, the corresponding edge variable $m_i$ selects either $u_i$ or $v_i$. Therefore, every edge has at least one
selected endpoint. Hence, the vertex set induced by any edge assignment covers all edges in $E$.

\textbf{Proposition II.}
Every minimum vertex cover $C^\star$ of $G$ can be induced by at least one edge assignment in the edge-centric representation.

\textit{Proof.}
For each edge $e_j=(u_j,v_j)$, since $C^\star$ is a vertex cover, at least one endpoint belongs to $C^\star$. If exactly one endpoint belongs to $C^\star$, we assign $m_j$ to select that endpoint; if both endpoints belong to $C^\star$, either endpoint may be selected. Thus, no vertex outside $C^\star$ is selected, and therefore
\[
C(\mathbf{m}) \subseteq C^\star.
\]

Now consider any vertex $u\in C^\star$. Because $C^\star$ is a minimum vertex cover, $u$ must have at least one neighbor $v\notin C^\star$; otherwise, removing $u$ would still leave all edges incident to $u$ covered, contradicting the minimality of $C^\star$. Hence there exists an edge $(u,v)$ with $u\in C^\star$ and $v\notin C^\star$. For this edge, the construction is forced to select $u$, so $u\in C(\mathbf{m})$. Therefore,
\[
C^\star \subseteq C(\mathbf{m}).
\]

Combining the two inclusions gives
\[
C(\mathbf{m}) = C^\star.
\]
Hence every minimum vertex cover is reachable by at least one edge assignment.

\subsubsection{Multiplicity of edge assignments} \label{multiplicity}

The Edge-Centric representation introduces multiplicity in the mapping from edge assignments to minimum vertex covers. For a given MVC, the corresponding edge assignment is generally not unique, multiple edge assignments may induce the same MVC.

Given a minimum vertex cover $C^\star$ of a graph $G=(V,E)$, consider an edge $e_i=(u_i,v_i)$. If exactly one endpoint belongs to $C^\star$, the edge assignment is uniquely determined. Specifically, if $u_i\in C^\star$ and $v_i\notin C^\star$, we set $m_i=0$ to select $u_i$, if $v_i\in C^\star$ and $u_i\notin C^\star$, we set $m_i=1$ to select $v_i$. We refer to such an edge as a \emph{crossing edge}.

In contrast, if both $u_i$ and $v_i$ belong to $C^\star$, the edge $e_i$ is an \emph{internal edge}. In this case, either value of $m_i$ may be used, since selecting either endpoint preserves the same induced vertex cover.

Let
\[
E_{in}(C^\star)
=
\{(u,v)\in E \mid u\in C^\star,\; v\in C^\star\}
\]
denote the set of internal edges of $C^\star$. Each internal edge provides two possible endpoint selections, whereas each crossing edge has only one valid selection. Therefore, the number of edge assignments that induce the same minimum vertex cover $C^\star$ is
\[
\mu(C^\star)
=
2^{|E_{in}(C^\star)|}.
\]

For example, consider a graph $G=(V,E)$ with
\[
V=\{0,1,2,3\}
\]
and
\[
E=\{(0,1),(1,2),(2,3),(0,3),(1,3)\}.
\]
The minimum vertex cover is
\[
C^\star=\{1,3\}.
\]

\begin{algorithm}[!t]
\caption{Edge-Centric Grover Search for MVC}
\label{alg:edge_centric}
\begin{algorithmic}[1]

\STATE \textbf{Input:} Graph $G=(V,E)$ with $|V|=n$ and $|E|=m$,
target cover size $k$, Grover iteration count $R_k$
\STATE \textbf{Output:} An edge assignment inducing a vertex cover
of size $k$

\STATE Allocate an $m$-qubit edge register $Q_E$
\STATE Allocate an $n$-qubit vertex register $Q_V$
\STATE Allocate Wallace-tree carry ancillas $Q_W$
\STATE Allocate a phase ancilla $q_p$ and initialize it to $\ket{-}$

\STATE Apply $H^{\otimes m}$ to $Q_E$ to prepare all edge assignments

\FOR{$r=1$ to $R_{k}$}

    \STATE \textit{// Vertex-state derivation}
    \FORALL{$v\in V$}
        \STATE Derive the state of $v$ from its incident edge variables
    \ENDFOR

    \STATE \textit{// Wallace-tree cardinality evaluation}
    \STATE Compute the Hamming weight of $Q_V$ using a reversible
    Wallace-tree reduction

    \STATE \textit{// Phase marking}
    \IF{the derived vertex cardinality equals $k$}
        \STATE Apply a phase flip using $q_p$
    \ENDIF

    \STATE \textit{// Uncomputation}
    \STATE Reverse the Wallace-tree computation
    \STATE Uncompute the derived vertex states

    \STATE \textit{// Edge-space diffusion}
    \STATE Apply the Grover diffusion operator to $Q_E$

\ENDFOR

\STATE Measure $Q_E$
\STATE Derive the corresponding vertex cover $C$
\RETURN $C$

\end{algorithmic}
\end{algorithm}

\begin{figure*}[t]
    \centerline{\includegraphics[width=1\textwidth]{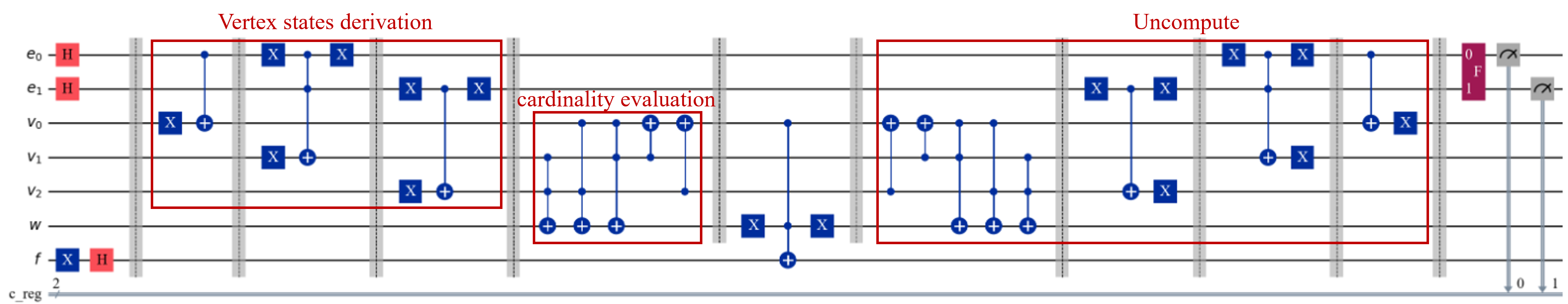}}
    \caption{Quantum circuit of the proposed Edge-Centric Grover search for MVC.}
    \label{fig:edgeEncircuit}
\end{figure*}

\begin{figure}[t]
    \centering

    \begin{subfigure}[b]{0.4\columnwidth}
        \centering
        \includegraphics[width=\linewidth]{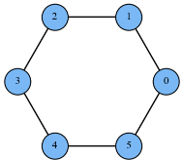}
        \caption{Sparse graph $C_6$.}
        \label{fig:sparse_graph}
    \end{subfigure}
    \hspace{0.03\columnwidth}
    \begin{subfigure}[b]{0.4\columnwidth}
        \centering
        \includegraphics[width=\linewidth]{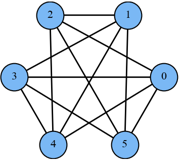}
        \caption{Dense graph $K_6-M$.}
        \label{fig:dense_graph}
    \end{subfigure}

    \caption{Representative graph instances from the sparse and dense
    benchmark families.}
    \label{fig:graph_families}
\end{figure}

For edge $(0,1)$, only vertex $1$ belongs to $C^\star$. Since vertex $1$ is the second endpoint, the corresponding edge variable must be set to $1$. Similarly, for edge $(1,2)$, only vertex $1$ belongs to $C^\star$, so the corresponding edge variable is $0$. For edges $(2,3)$ and $(0,3)$, only vertex $3$ belongs to $C^\star$, and hence their edge variables are both set to $1$.

The remaining edge $(1,3)$ is an internal edge because both endpoints belong to $C^\star$. Therefore, its edge variable may be assigned either $0$ or $1$ without changing the induced vertex cover.

Thus, under the edge ordering
\[
(0,1),(1,2),(2,3),(0,3),(1,3),
\]
the two edge assignments
\[
10110
\quad\text{and}\quad
10111
\]
both induce the same minimum vertex cover
\[
C^\star=\{1,3\}.
\]
Hence, the multiplicity of $C^\star$ is
\[
\mu(C^\star)=2^{|E_{in}(C^\star)|}=2^1=2.
\]

For a graph with multiple minimum vertex covers, each MVC may have a different number of internal edges and therefore a different multiplicity. Let $\mathcal{C}^\star$ denote the set of all minimum vertex covers. The total number of edge assignments that induce a minimum vertex cover is then
\[
M_{edge} = \sum_{C\in\mathcal{C}^\star} 2^{|E_{in}(C)|},
\]
where $E_{in}(C)$ is the set of internal edges associated with minimum vertex cover $C$.

\subsubsection{Implementation}

Algorithm~\ref{alg:edge_centric} summarizes the Edge-Centric Grover search for MVC. First, Hadamard gates are applied to the $m$-qubit edge register $Q_E$ to prepare an equal superposition over all possible edge assignments. The corresponding vertex-selection states are then derived from the edge variables as described in Subsection~\ref{vetex_derivation}.

The cardinality of the derived vertex set is evaluated using a reversible tree-based population-count circuit inspired by the carry-save reduction structure of a Wallace tree. At each reduction level, groups of three input bits are compressed into sum and carry bits using reversible full-adder blocks, following the parallel 3-to-2 reduction principle used in reversible Wallace-tree arithmetic architectures~\cite{wallace,rabaey}. Rather than accumulating the vertex-selection bits sequentially, the $n$ derived vertex states are treated as parallel binary inputs and reduced through hierarchical carry-save compression. At each level, groups of three signals are processed by reversible full adders to produce sum and carry outputs, implementing a 3-to-2 reduction. These operations can be performed in parallel across disjoint groups, and the resulting partial sums and carries are recursively reduced until the Hamming weight of the derived vertex set is obtained.

The oracle then applies a phase flip to edge assignments whose induced vertex cover has cardinality $k$. After uncomputing the Wallace-tree and vertex-derivation circuits, the Grover diffusion operator is applied to the edge register to amplify the amplitudes of the marked edge assignments. These steps are repeated for the required number of Grover iterations before the edge register is measured and the corresponding vertex cover is recovered.

As illustrated in Figure~\ref{fig:edgeEncircuit}, consider a simple graph with edges $(0,1)$ and $(1,2)$. Two Hadamard gates are first applied to the edge qubits $m_0$ and $m_1$ to prepare an equal superposition over all possible edge assignments. The vertex-derivation circuit then computes the corresponding states of the vertex qubits according to the edge-qubit assignments. A Wallace-style reversible tree adder is subsequently applied to evaluate the Hamming weight of the derived vertex set. Since the minimum vertex cover of this graph has cardinality one, corresponding to vertex $1$, the oracle marks the states for which the cardinality output equals the binary value $01$. After the cardinality-evaluation and vertex-derivation circuits are uncomputed, the Grover diffusion operator is applied to the edge register to amplify the amplitudes of the marked edge assignments. 


\begin{figure}[t]
    \centering

    \begin{subfigure}[b]{0.4\columnwidth}
        \centering
        \includegraphics[width=\linewidth]{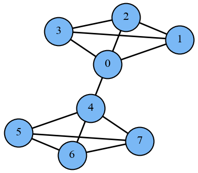}
        \caption{$m=13$, $\rho=46.4\%$}
        \label{fig:density46}
    \end{subfigure}
    \hspace{0.03\columnwidth}
    \begin{subfigure}[b]{0.4\columnwidth}
        \centering
        \includegraphics[width=\linewidth]{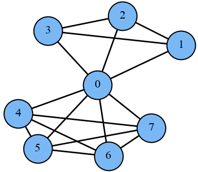}
        \caption{$m=16$, $\rho=57.1\%$}
        \label{fig:density57}
    \end{subfigure}

    \vspace{0.4em}

    \begin{subfigure}[b]{0.4\columnwidth}
        \centering
        \includegraphics[width=\linewidth]{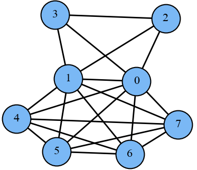}
        \caption{$m=20$, $\rho=71.4\%$}
        \label{fig:density71}
    \end{subfigure}
    \hspace{0.03\columnwidth}
    \begin{subfigure}[b]{0.4\columnwidth}
        \centering
        \includegraphics[width=\linewidth]{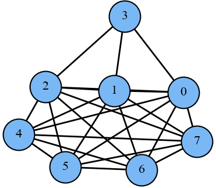}
        \caption{$m=24$, $\rho=85.7\%$}
        \label{fig:density85}
    \end{subfigure}

    \caption{Fixed-$n$ density-sensitivity graph instances with
    $n=8$ and $k=6$.}
    \label{fig:density_graphs}
    \vspace{-0.3in}
\end{figure}
\section{Evaluation} \label{evaluation}

\subsection{Experimental Setup} \label{setup}



\begin{figure}[t]
    \centering

    \begin{subfigure}[t]{0.24\textwidth}
        \centering
        \includegraphics[width=\linewidth]{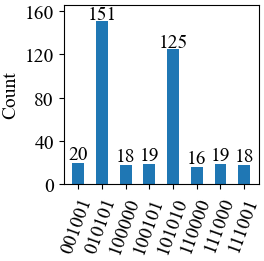}
        \caption{Baseline}
        \label{fig:s_baseline}
    \end{subfigure}
    \hfill
    \begin{subfigure}[t]{0.24\textwidth}
        \centering
        \includegraphics[width=\linewidth]{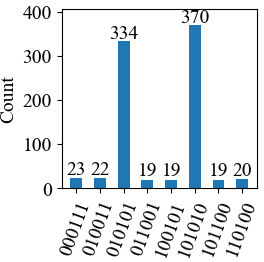}
        \caption{Dicke-Parallel}
        \label{fig:s_dicke}
    \end{subfigure}

    \vspace{0.1em}

    \begin{subfigure}[t]{0.24\textwidth}
        \centering
        \includegraphics[width=\linewidth]{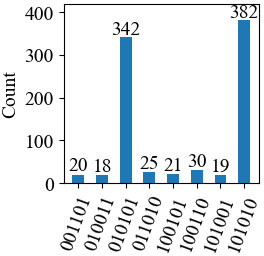}
        \caption{Edge-Counting}
        \label{fig:s_counting}
    \end{subfigure}
    \hfill
    \begin{subfigure}[t]{0.24\textwidth}
        \centering
        \includegraphics[width=\linewidth]{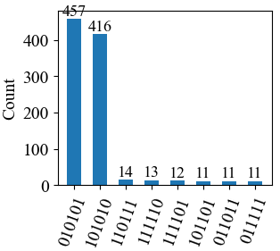}
        \caption{Edge-Centric}
        \label{fig:s_centric}
    \end{subfigure}

    \caption{Functional validation on the representative sparse graph.
   }
    \label{fig:functional_sparse}
\end{figure}

\begin{figure}[t]
    \centering

    \begin{subfigure}[t]{0.24\textwidth}
        \centering
        \includegraphics[width=\linewidth]{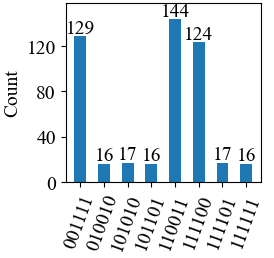}
        \caption{Baseline}
        \label{fig:d_baseline}
    \end{subfigure}
    \hfill
    \begin{subfigure}[t]{0.24\textwidth}
        \centering
        \includegraphics[width=\linewidth]{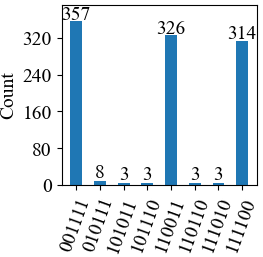}
        \caption{Dicke-Parallel}
        \label{fig:d_dicke}
    \end{subfigure}

    \vspace{0.1em}

    \begin{subfigure}[t]{0.24\textwidth}
        \centering
        \includegraphics[width=\linewidth]{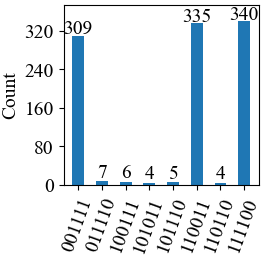}
        \caption{Edge-Counting}
        \label{fig:d_counting}
    \end{subfigure}
    \hfill
    \begin{subfigure}[t]{0.24\textwidth}
        \centering
        \includegraphics[width=\linewidth]{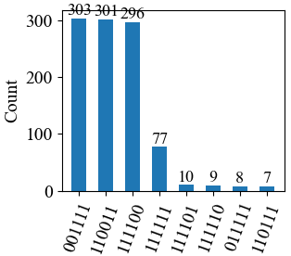}
        \caption{Edge-Centric}
        \label{fig:d_centric}
    \end{subfigure}

    \caption{Functional validation on the representative dense graph.
    }
    \label{fig:functional_dense}
\end{figure}

We implemented all proposed approaches using the Qiskit framework~\cite{qiskit}. 
Circuit correctness was evaluated using the Qiskit Aer Sampler with 1024 shots. 
For a consistent logical-resource comparison, all multi-controlled X (MCX) operations were synthesized using the no-ancilla Huang--Palsberg MCX synthesis~\cite{huang}. The resulting circuits were then decomposed into the common gate basis $\{U,{CX}\}$ with transpiler optimization 
disabled (\texttt{optimization\_level=0}). No hardware-specific mapping or routing was applied. 
We report the algorithmic qubit count prior to gate decomposition, together with the decomposed logical depth, total logical gate count, and CX gate count. Since CX is the only two-qubit gate in the selected basis, the CX count is also reported as the logical two-qubit gate count (2Q gate count).

To examine the impact of graph structure, we evaluate the proposed approaches using three groups of graph instances. The sparse benchmarks are cycle graphs $C_n$, for which
$m=n$ and the graph density is
\[
\rho_{{sparse}}=\frac{2}{n-1}.
\]
The dense benchmarks are constructed as $K_n-M$, where $M$ is a
perfect matching removed from the complete graph $K_n$. Specifically,
$M$ consists of $n/2$ pairwise disjoint edges,
\[
M=\{(0,1),(2,3),\ldots,(n-2,n-1)\},
\]
such that every vertex belongs to exactly one removed edge. Therefore,
the resulting graph remains highly connected while excluding one
incident edge for each vertex. These graphs contain
\[
m=\frac{n(n-2)}{2}
\]
edges and have density
\[
\rho_{{dense}}=\frac{n-2}{n-1}.
\]

To isolate the effect of graph density from graph size, we additionally construct a fixed-$n$ graph family with $n=8$. The vertices are divided into two groups,
\[
A=\{0,1,2,3\}, \qquad B=\{4,5,6,7\},
\]
with each group forming a $K_4$ clique. Starting from these two cliques, cross edges between $A$ and $B$ are progressively added to increase the graph density while keeping the number of vertices fixed. We evaluate four configurations containing $1$, $4$, $8$, and $12$ cross edges, corresponding to $m=13$, $16$, $20$, and $24$ total edges and graph densities of $46.4\%$, $57.1\%$, $71.4\%$, and $85.7\%$, respectively.

The graphs are constructed as a nested family such that each denser instance contains all edges of the preceding instance. For all four configurations, the maximum independent-set size remains two, and hence the minimum vertex cover size remains fixed at $k=6$. This setup therefore allows us to examine the impact of increasing edge density without simultaneously changing the number of vertices or the target MVC cardinality.

Representative examples of the sparse and dense graph families are shown in Figure~\ref{fig:graph_families}. The four fixed-$n$ density-sensitivity instances are shown separately in Figure~\ref{fig:density_graphs} to illustrate the progressive addition of cross edges as graph density increases.

These graph families provide representative test cases for functional validation and backend-independent logical-resource comparison in terms of algorithmic qubit count, decomposed logical depth, total logical gate count, and logical two-qubit (CX) gate count.

\subsection{Functional Validation}

\begin{figure}[t]
    \centering

    \begin{subfigure}[t]{0.48\columnwidth}
        \centering
        \includegraphics[width=\linewidth]{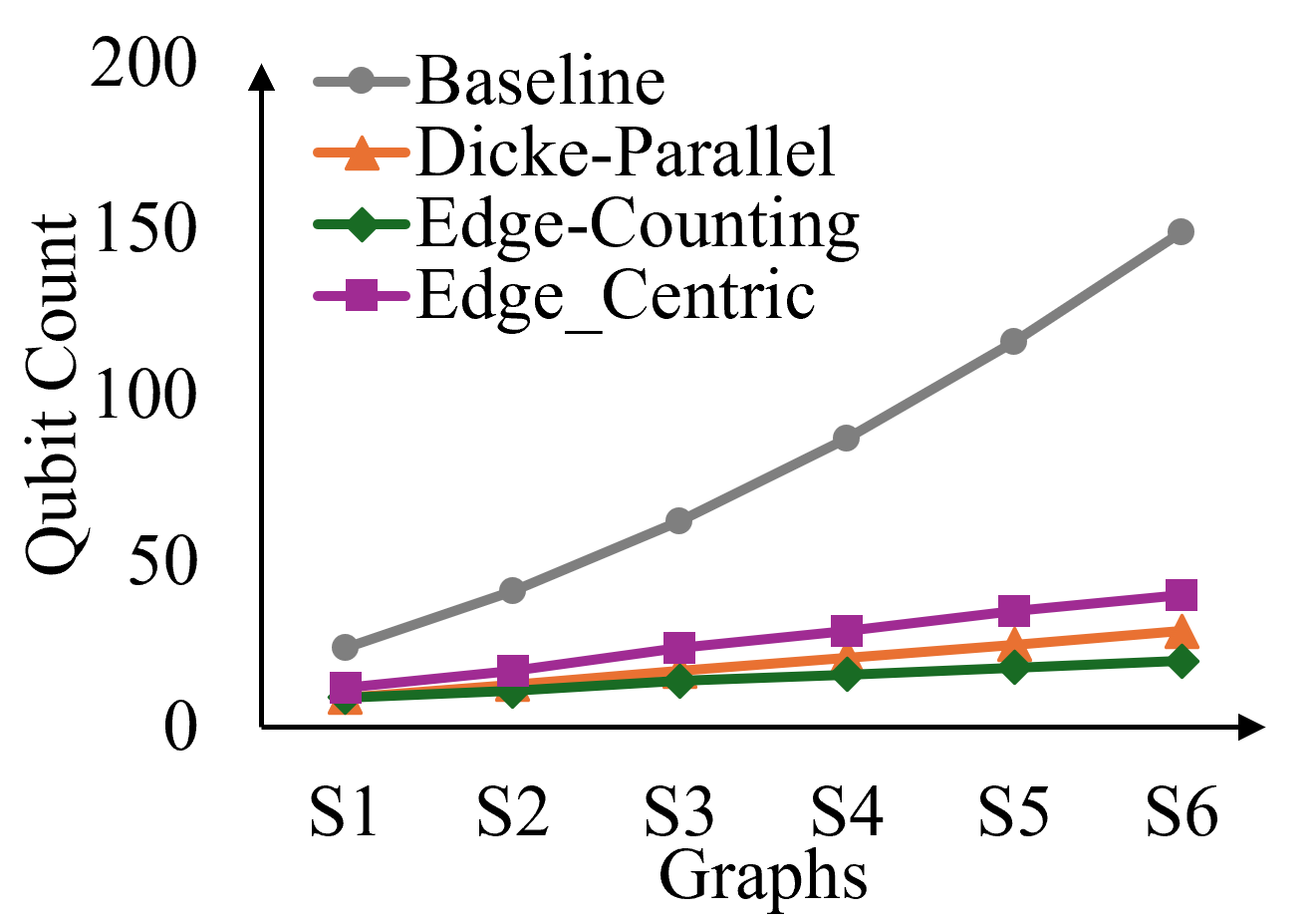}
        \caption{Qubit count}
        \label{squbits}
    \end{subfigure}
    \hfill
    \begin{subfigure}[t]{0.48\columnwidth}
        \centering
        \includegraphics[width=\linewidth]{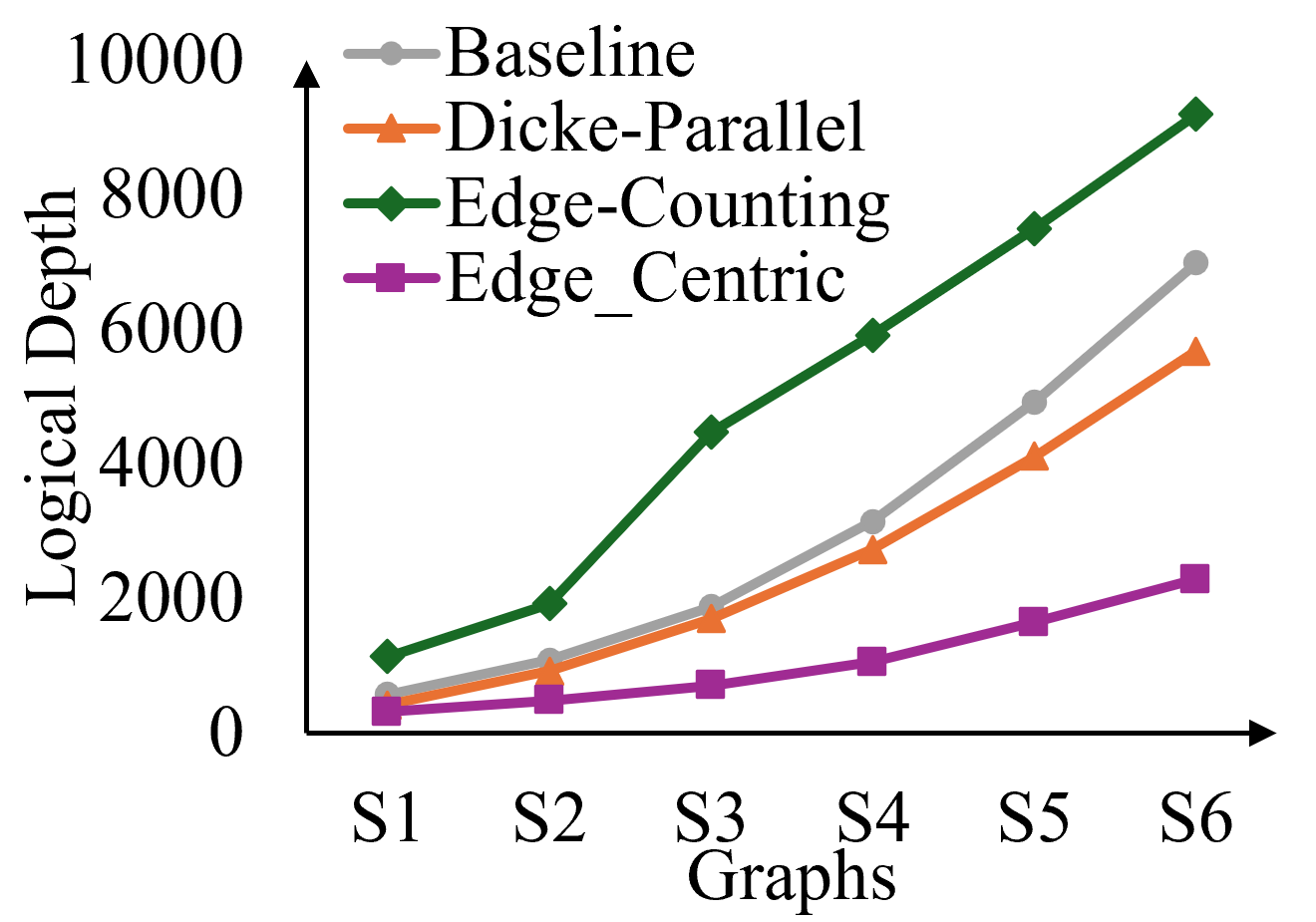}
        \caption{Depth}
        \label{sdepth}
    \end{subfigure}

    \vspace{0.4em}

    \begin{subfigure}[t]{0.48\columnwidth}
        \centering
        \includegraphics[width=\linewidth]{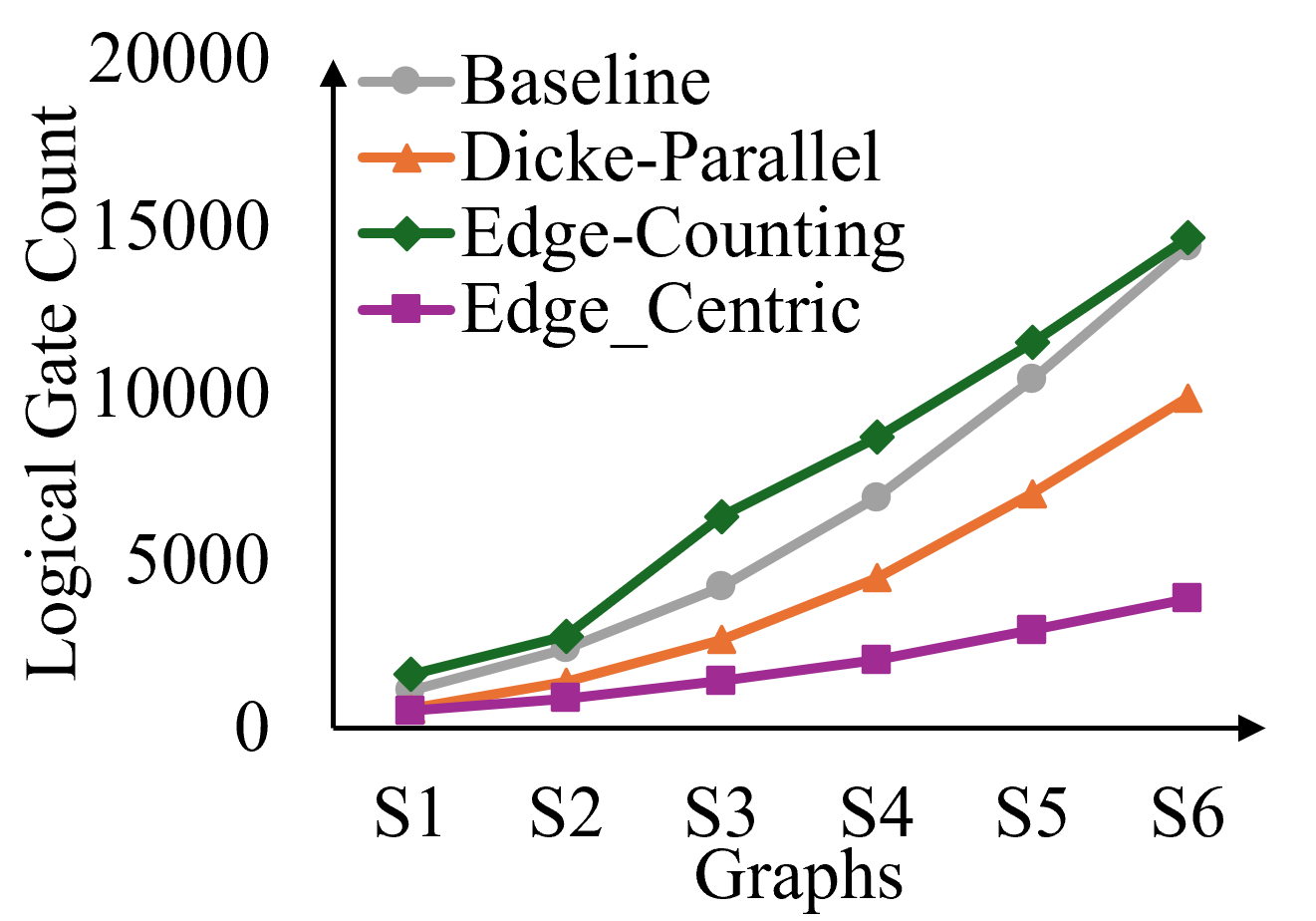}
        \caption{Gate count}
        \label{sgates}
    \end{subfigure}
    \hfill
    \begin{subfigure}[t]{0.48\columnwidth}
        \centering
        \includegraphics[width=\linewidth]{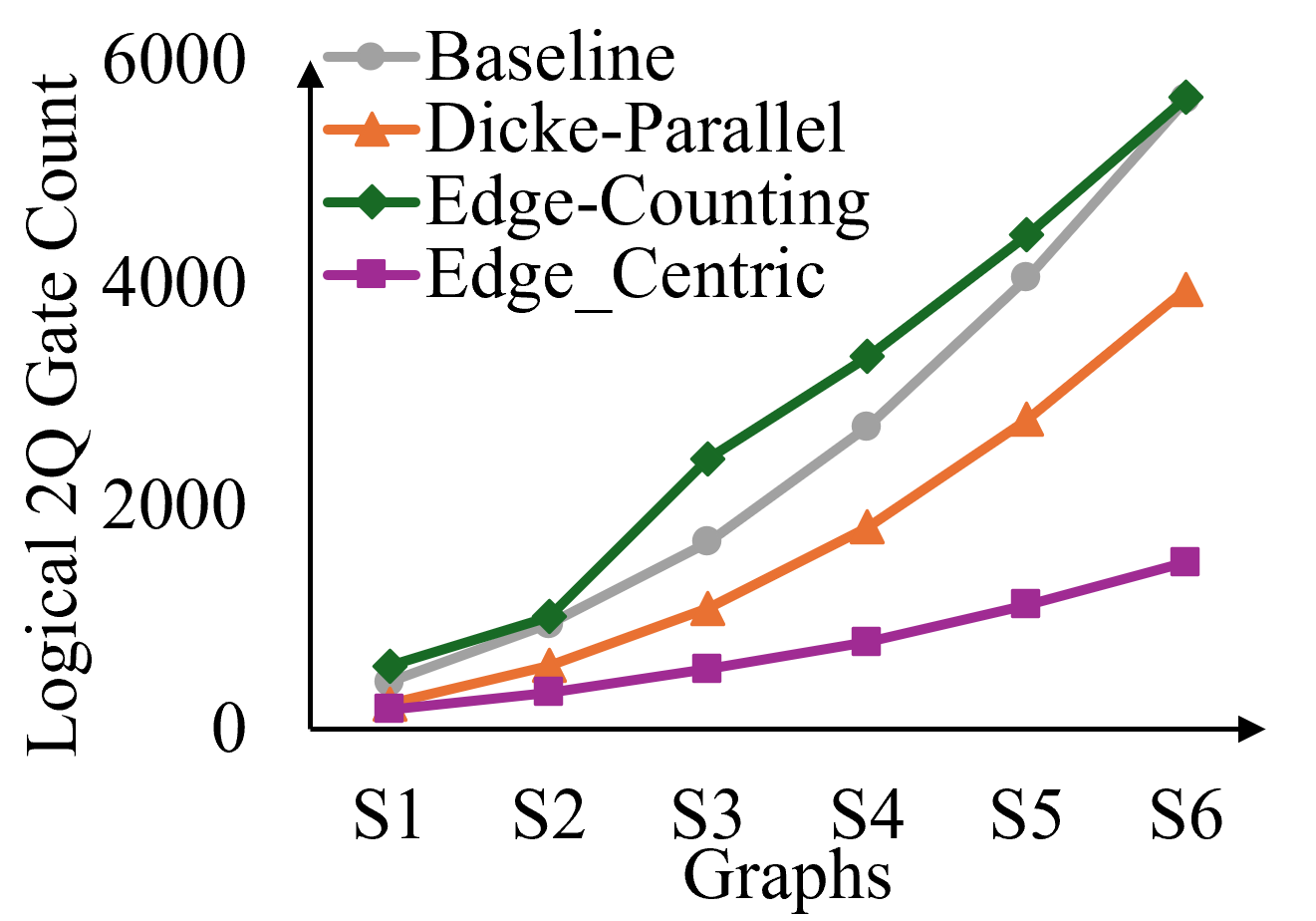}
        \caption{2Q gate count}
        \label{s2q}
    \end{subfigure}

    \caption{Resource comparison on sparse graphs. }
    \label{fig:sresource-comparison}
    \vspace{-0.2in}
\end{figure}

Functional validation is conducted on the representative graph instances shown in Figure~\ref{fig:graph_families}, covering both sparse and dense graph structures. For each instance, the quantum-circuit output is compared with the classically determined minimum vertex covers to verify functional correctness. Larger graph instances are reserved for resource evaluation, as full circuit simulation becomes increasingly expensive with growing circuit size.

Figure~\ref{fig:functional_sparse} shows the results for the representative sparse graph in Figure~\ref{fig:sparse_graph}. For clarity, we plot the top-8 results. As shown, all four
implementations amplify the measurement outcomes corresponding to the two valid minimum vertex covers,
$C=\{0,2,4\}$ and $C=\{1,3,5\}$.

Figure~\ref{fig:functional_dense} shows the results for the representative dense graph in Figure~\ref{fig:dense_graph}. All four implementations successfully amplify the minimum vertex cover solutions,
$C=\{0,1,2,3\}$, $C=\{0,1,4,5\}$, and $C=\{2,3,4,5\}$.
For the Edge-Centric design, the measured edge assignments are decoded and aggregated according to their corresponding vertex-selection states before visualization.

\subsection{Resource Analysis}

\begin{figure}[t]
    \centering

    \begin{subfigure}[t]{0.48\columnwidth}
        \centering
        \includegraphics[width=\linewidth]{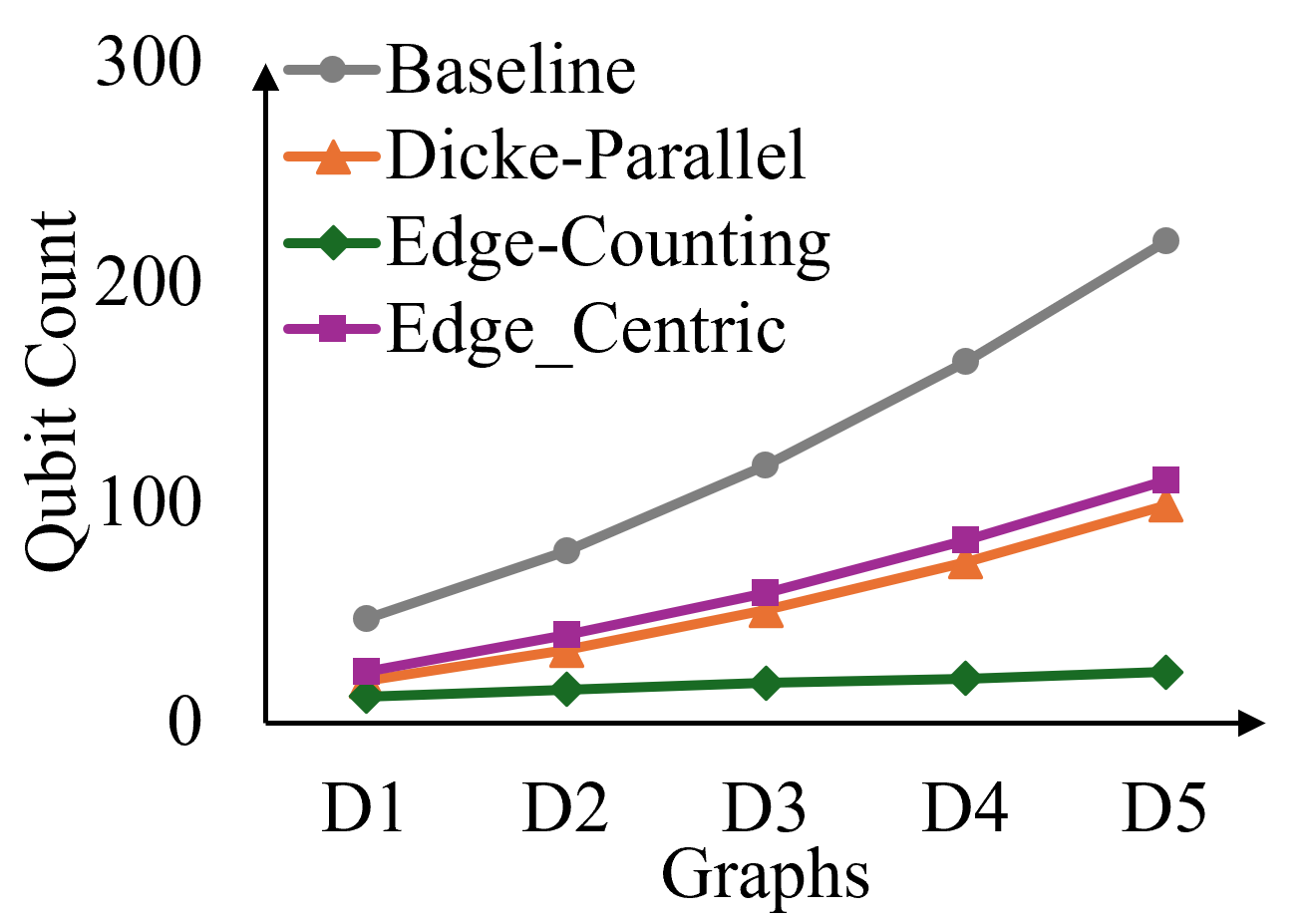}
        \caption{Qubit count}
        \label{dqubits}
    \end{subfigure}
    \hfill
    \begin{subfigure}[t]{0.48\columnwidth}
        \centering
        \includegraphics[width=\linewidth]{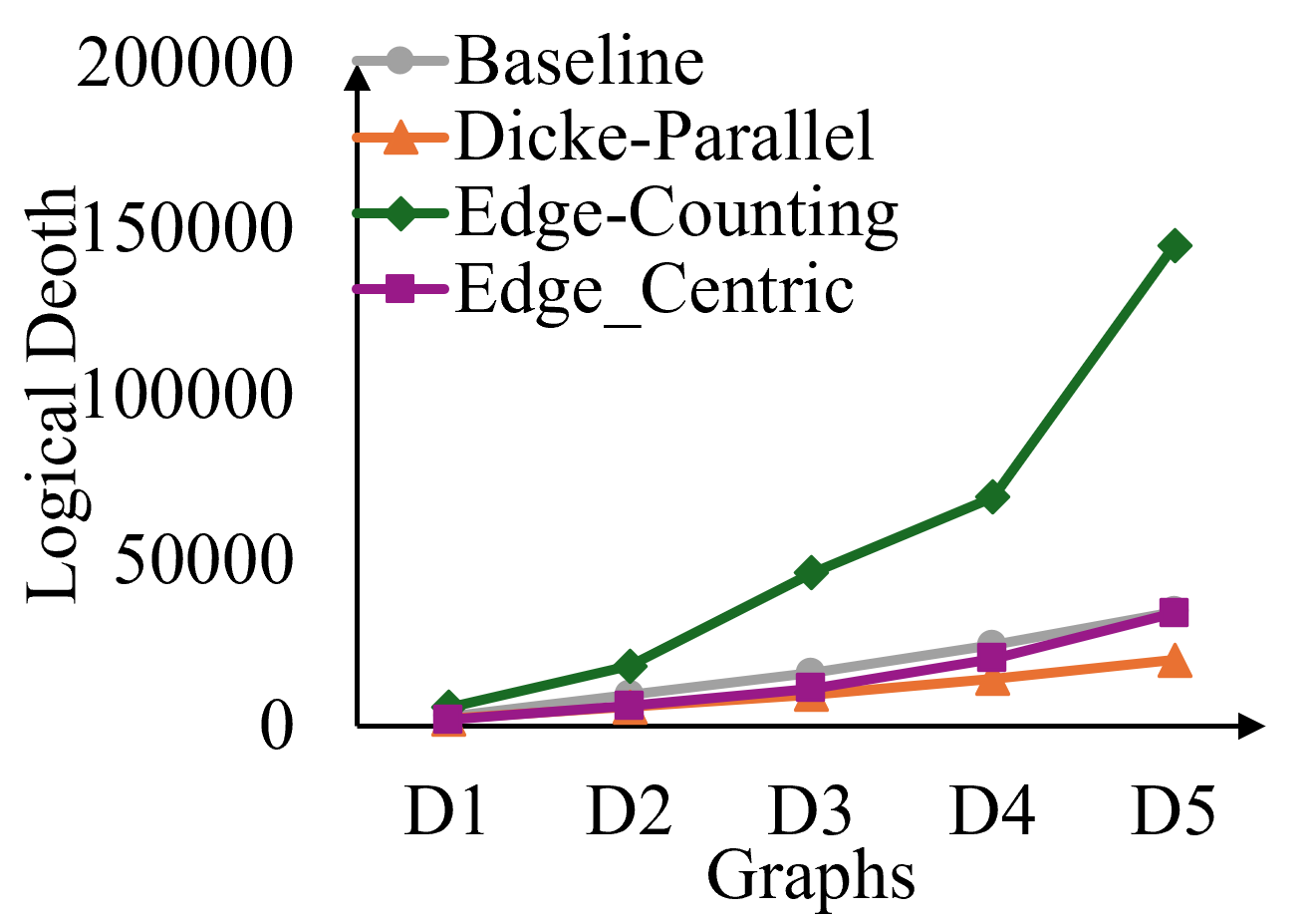}
        \caption{Depth}
        \label{ddepth}
    \end{subfigure}

    \vspace{0.4em}

    \begin{subfigure}[t]{0.48\columnwidth}
        \centering
        \includegraphics[width=\linewidth]{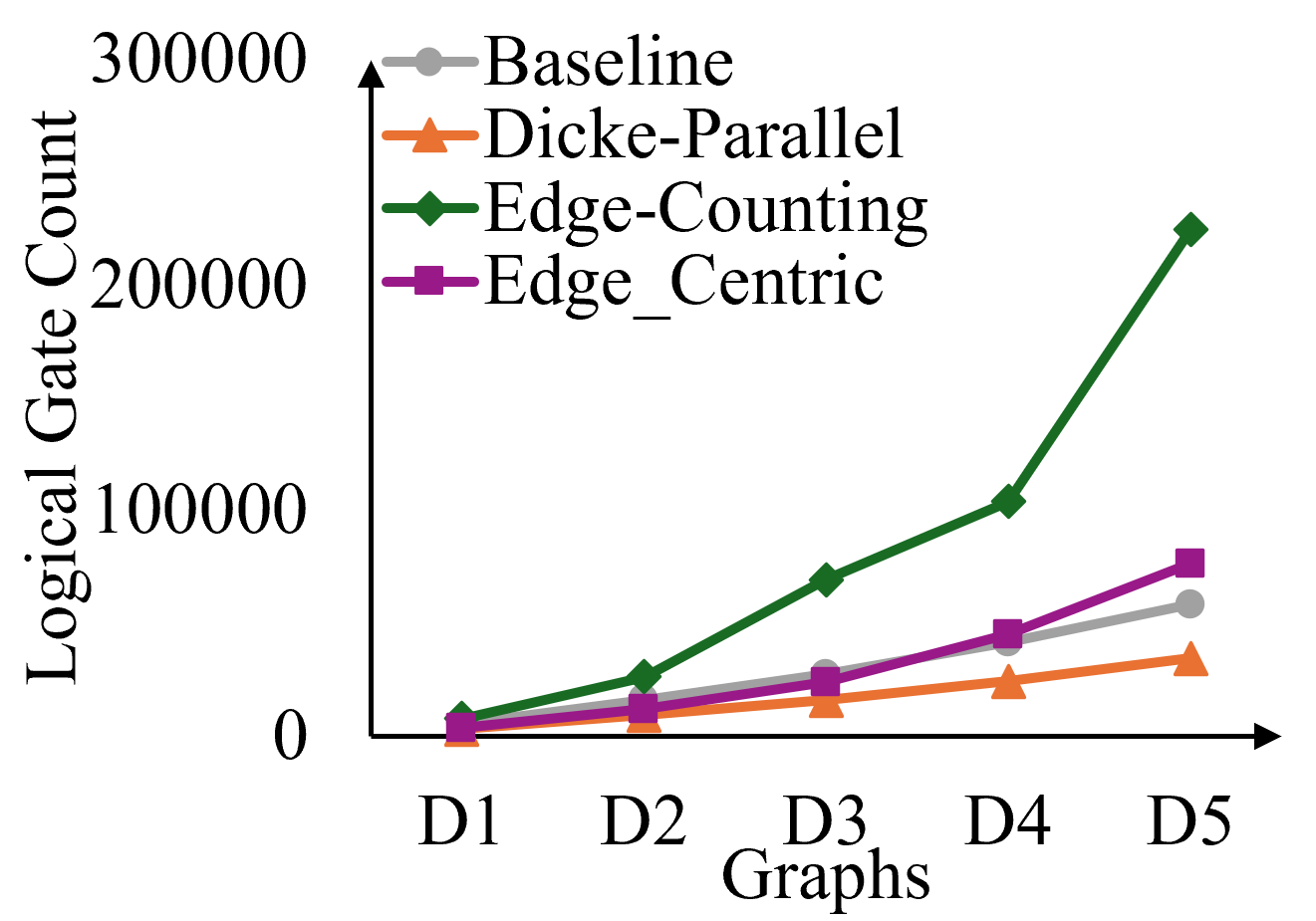}
        \caption{Gate count}
        \label{dgates}
    \end{subfigure}
    \hfill
    \begin{subfigure}[t]{0.48\columnwidth}
        \centering
        \includegraphics[width=\linewidth]{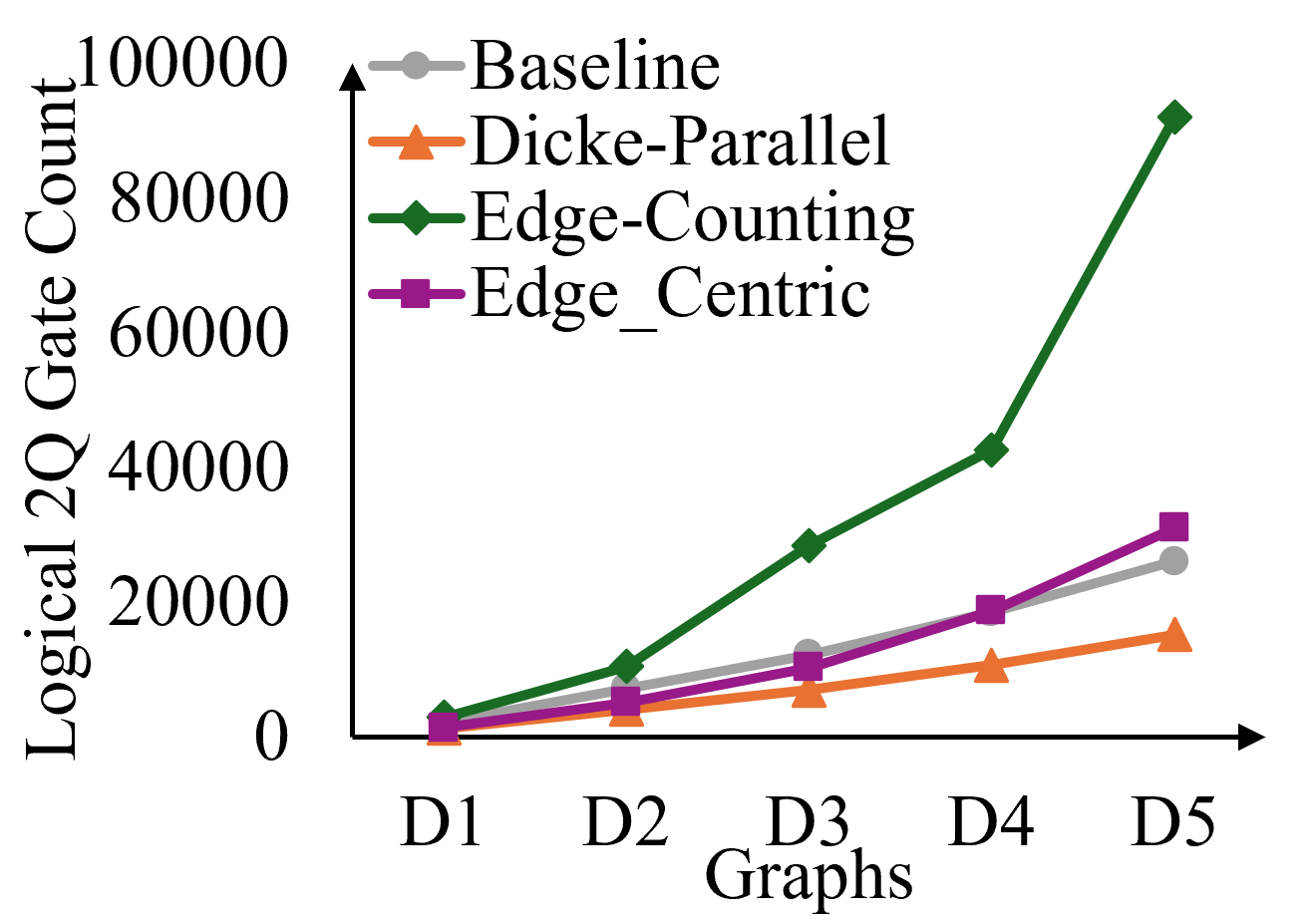}
        \caption{2Q gate count}
        \label{d2q}
    \end{subfigure}

    \caption{Resource comparison on dense graphs.}
    \label{fig:dresource-comparison}
\end{figure}

\begin{figure}[t]
    \centering

    \begin{subfigure}[t]{0.48\columnwidth}
        \centering
        \includegraphics[width=\linewidth]{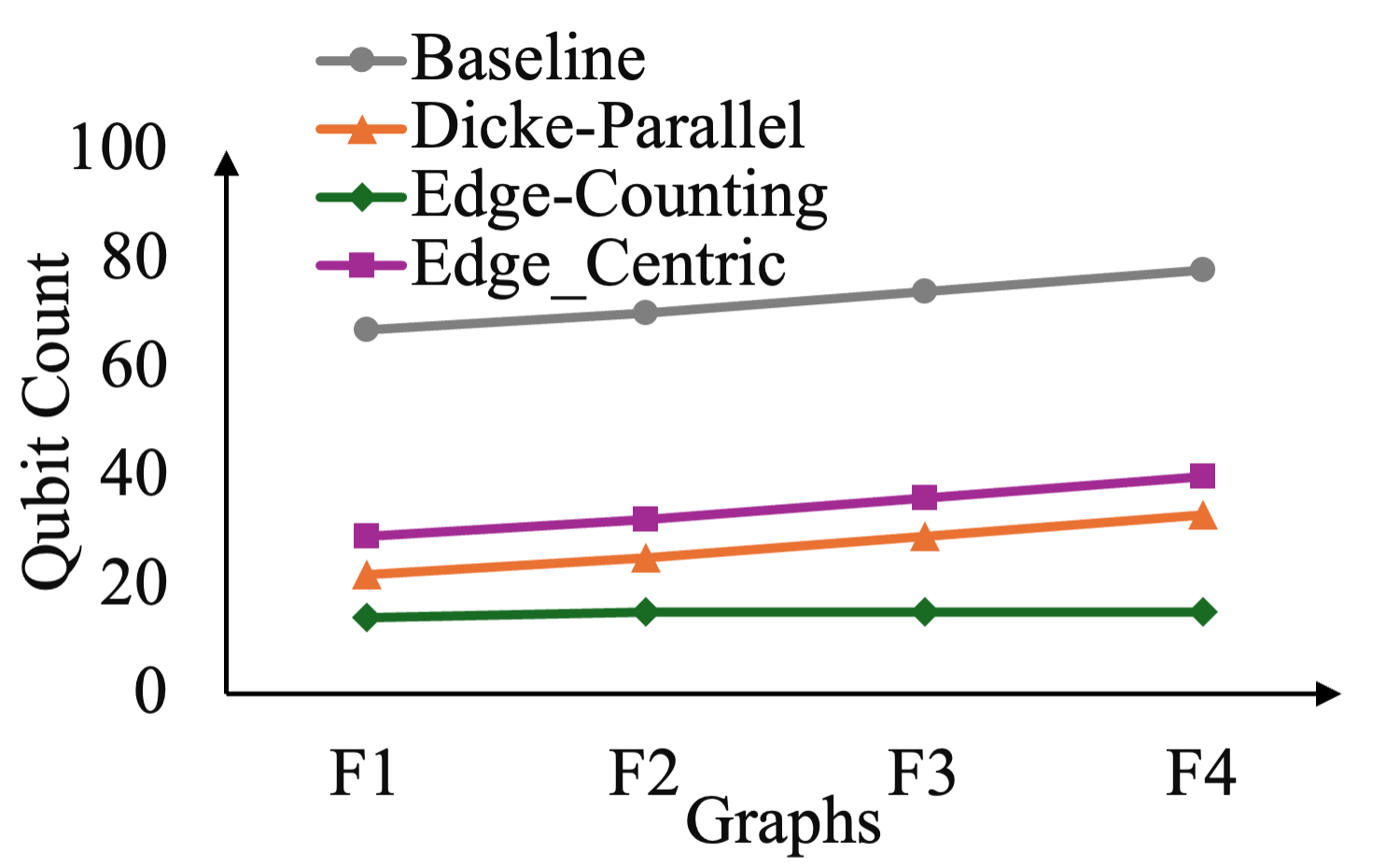}
        \caption{Qubit count}
        \label{fqubits}
    \end{subfigure}
    \hfill
    \begin{subfigure}[t]{0.48\columnwidth}
        \centering
        \includegraphics[width=\linewidth]{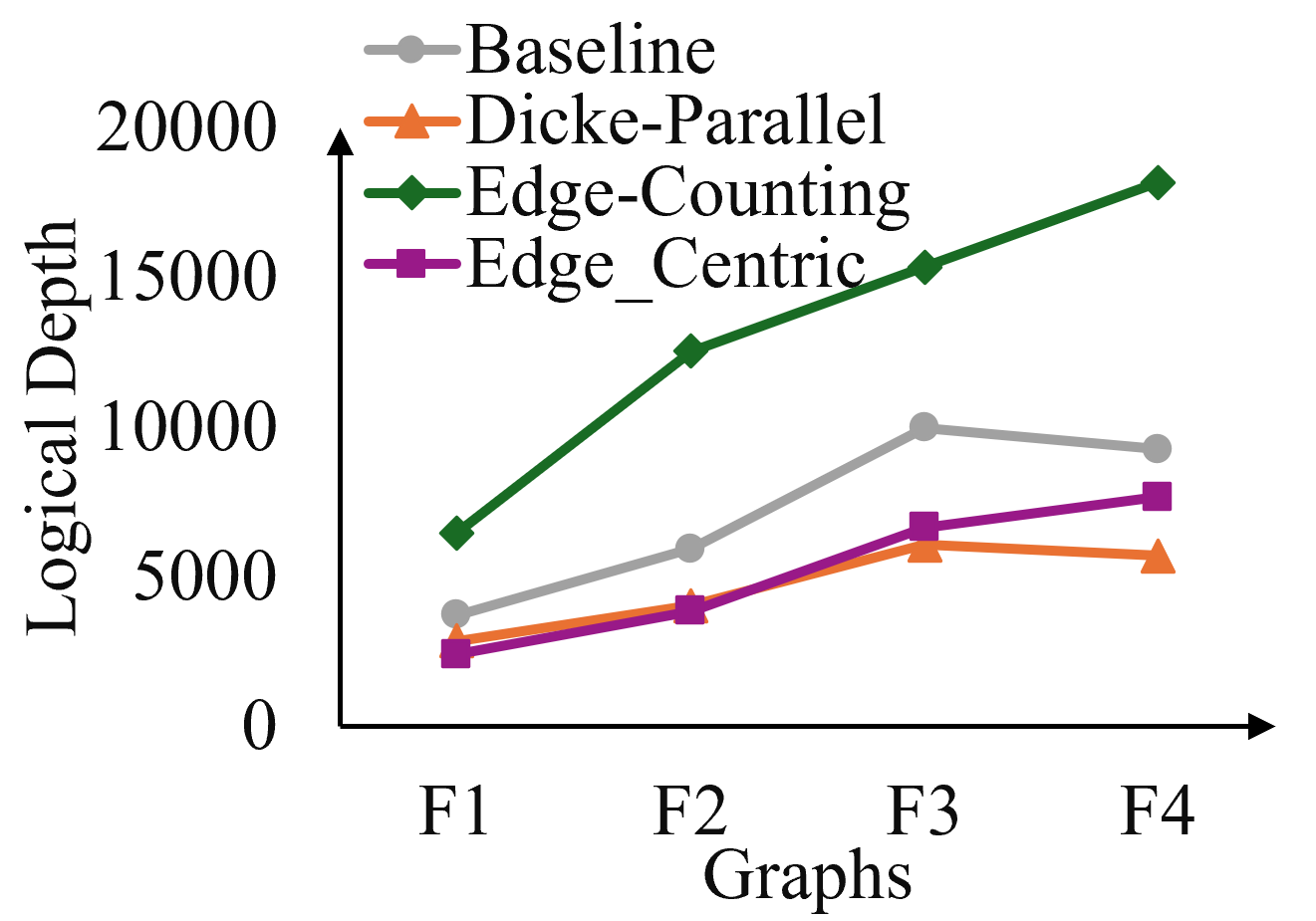}
        \caption{Depth}
        \label{fdepth}
    \end{subfigure}

    \vspace{0.4em}

    \begin{subfigure}[t]{0.48\columnwidth}
        \centering
        \includegraphics[width=\linewidth]{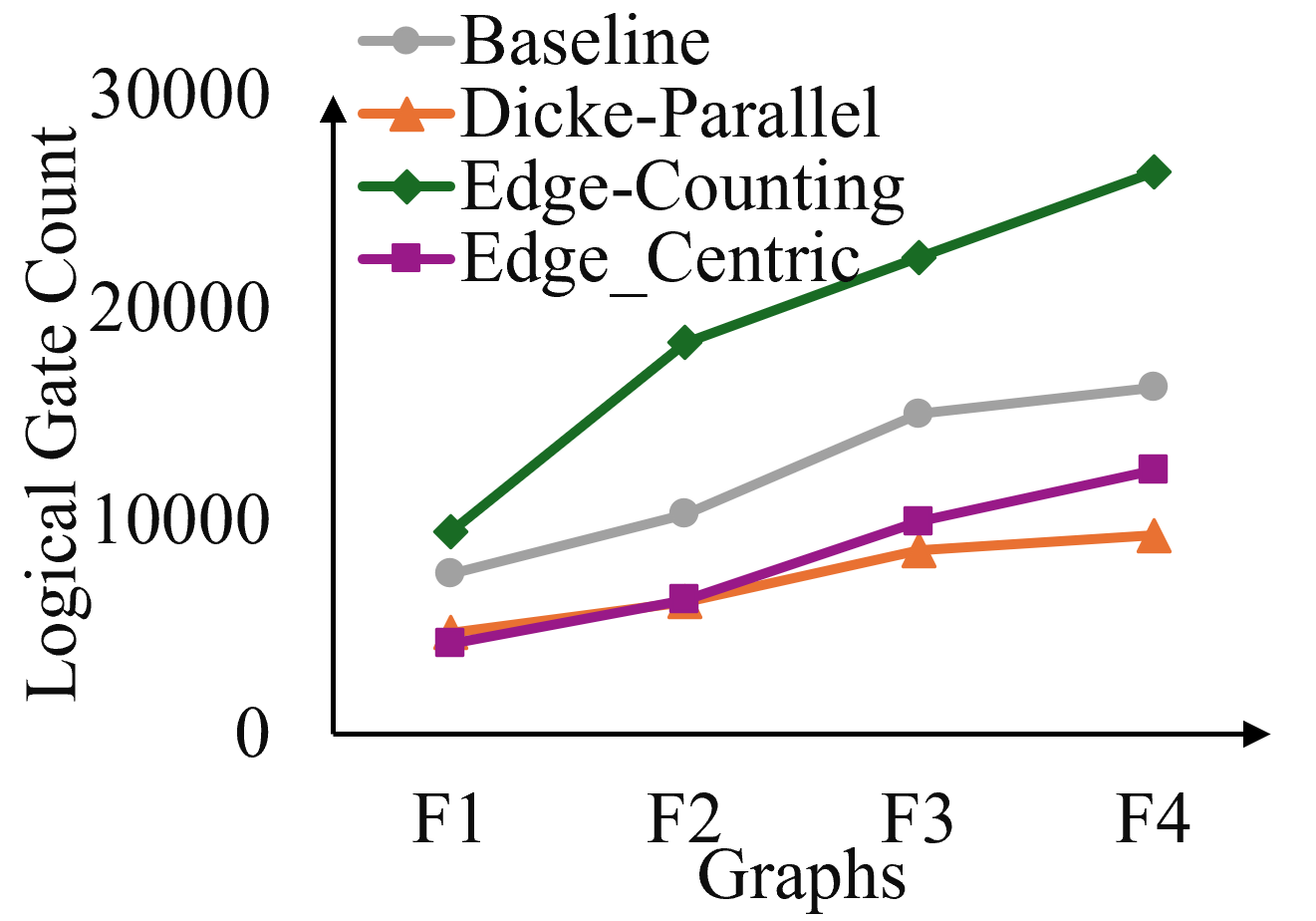}
        \caption{Gate count}
        \label{fgates}
    \end{subfigure}
    \hfill
    \begin{subfigure}[t]{0.48\columnwidth}
        \centering
        \includegraphics[width=\linewidth]{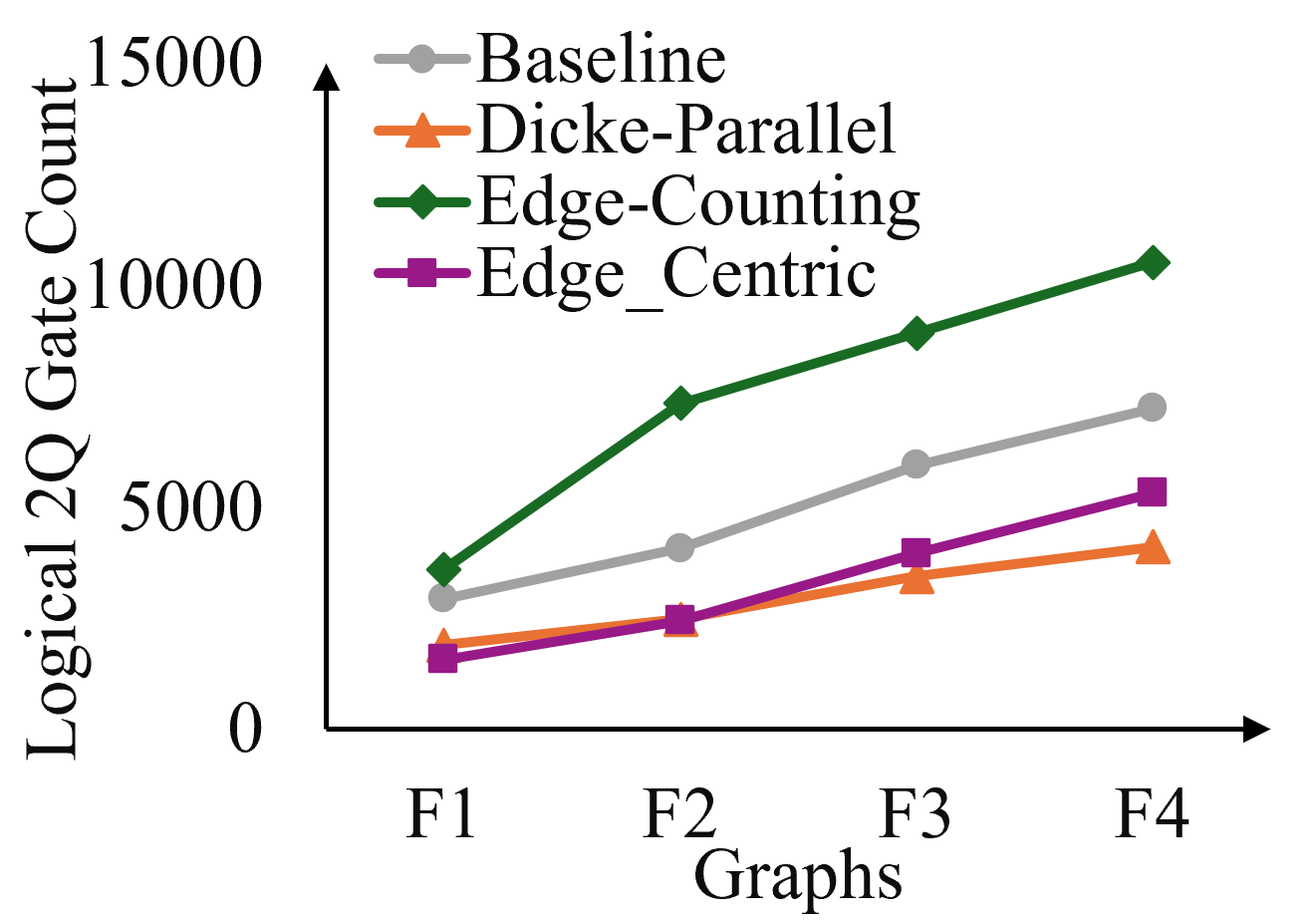}
        \caption{2Q gate count}
        \label{f2q}
    \end{subfigure}

    \caption{Resource comparison on fix-$n$ graphs with increasing density.}
    \label{fig:fresource-comparison}
    \vspace{-0.2in}
\end{figure}

In this subsection, we evaluate the resource efficiency of the proposed Grover-search designs. The results are presented in
Figures~\ref{fig:sresource-comparison},
\ref{fig:dresource-comparison}, and
\ref{fig:fresource-comparison}.

\begin{figure}[t]
    \centering
    \includegraphics[width=0.88\columnwidth]{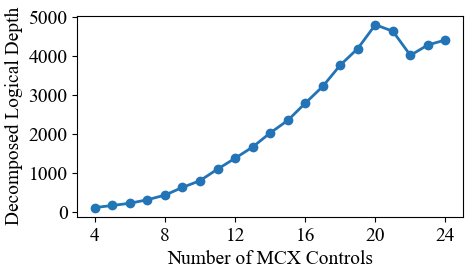}
    \caption{Decomposed logical depth of the Huang--Palsberg no-ancilla MCX synthesis as a function of the number of control qubits~\cite{huang}.}
    \label{fig:mcx_depth}
    \vspace{-0.1in}
\end{figure}

\subsubsection{One Grover Iteration}

We first evaluate the quantum resources required to execute a single Grover iteration. Figure~\ref{fig:sresource-comparison} compares the different MVC implementations on the sparse benchmark graphs described in Section~\ref{setup}. These benchmarks are cycle graphs, for which the number of edges is equal to the number of vertices, i.e., $|E|=|V|=n$. The smallest instance, $S1$, contains $n=4$ vertices, and the graph size increases by two vertices for each subsequent instance, reaching $n=14$ for $S6$.

As shown in Figure~\ref{fig:sresource-comparison}, Edge-Counting requires the fewest qubits across the sparse benchmarks. In contrast, the Edge-Centric architecture achieves substantially lower decomposed logical depth, total logical gate count, and two-qubit gate count. Overall, Edge-Centric provides the best performance for these three metrics, while Edge-Counting remains the most qubit-efficient architecture.

We then evaluate the resource efficiency of the proposed Grover-based architectures for MVC on the dense benchmark graphs described in Section~\ref{setup}. The results are shown in Figure~\ref{fig:dresource-comparison}. Each dense benchmark is constructed as $G = K_n - M$, where $M$ is a perfect matching of the complete graph $K_n$. Therefore, the number of edges is

\[
m = |E| = \frac{n(n-1)}{2} - \frac{n}{2}.
\]

$D1$ contains $n=6$ vertices, $D2$ contains $n=8$, and the graph size increases by two vertices for each subsequent instance.

For the dense benchmarks, Dicke-Parallel achieves the lowest decomposed logical depth, total logical gate count, and two-qubit gate count. Edge-Counting, however, again requires the fewest qubits and therefore remains the most qubit-efficient architecture.

This trend is consistent with the characteristics of the dense benchmarks. As edge density increases, both Dicke-Parallel and Edge-Centric require additional circuit resources. For Dicke-Parallel, each additional edge introduces an additional Toffoli-based operation for determining the corresponding edge status. In contrast, in the Edge-Centric design, each edge state must contribute to the derived states of both endpoint vertices, resulting in two edge-to-vertex update operations per edge. Consequently, as the number of edges increases, the edge-to-vertex derivation stage of Edge-Centric grows faster than the corresponding edge-status computation in Dicke-Parallel.

We next evaluate the impact of graph density while holding the number of vertices fixed, using the density-sensitivity graph family shown in Figure~\ref{fig:density_graphs}. The corresponding resource results are presented in Figure~\ref{fig:fresource-comparison}. The four instances contain 13, 16, 20, and 24 edges for $F1$ through $F4$, respectively.

As shown in Figure~\ref{fig:fresource-comparison}, Edge-Counting remains the most qubit-efficient design across all density levels. For decomposed logical depth, total logical gate count, and CX gate count, Edge-Centric is more resource-efficient for sparse and moderately dense graphs.
However, as graph density increases, Dicke-Parallel becomes increasingly competitive and eventually requires fewer resources for these metrics.

A local decrease in decomposed logical depth is observed from $F3$ to $F4$ for both Dicke-Parallel and the Baseline. This behavior originates from the no-ancilla Huang--Palsberg MCX synthesis~\cite{huang} used in our resource evaluation. The synthesized MCX depth is not strictly monotonic with the number of control qubits, as independently characterized in Figure~\ref{fig:mcx_depth}. In particular, the 20-control MCX has a decomposed depth of 4797, whereas the 24-control
MCX has a smaller depth of 4405, despite requiring more CX gates.
Since $F3$ contains 20 edges and therefore introduces a 20-control edge-aggregation MCX, while $F4$ introduces the corresponding 24-control MCX, this synthesis-level non monotonicity produces the local reduction in whole-circuit depth. Thus, the decrease should be interpreted as a property of the MCX synthesis and its critical path, rather than as a reduction in the underlying MVC workload.

\subsubsection{Number of Grover Iterations}

The circuit-resource analysis in the previous subsection evaluates the cost of a single Grover iteration. In this subsection, we focus only on the number of amplitude-amplification iterations required by the different MVC formulations.

For a search space containing $N$ basis states, among which $M$ states are marked, the number of Grover iterations required to approach the maximum success probability is approximately
\[
R(N,M)\approx \frac{\pi}{4}\sqrt{\frac{N}{M}}.
\]

Therefore, the formulation with the larger marked-state fraction requires fewer Grover iterations~\cite{Giter}.

\paragraph{Baseline Grover Search for MVC}
The conventional vertex-based formulation searches over all possible vertex subsets. For a graph with $n$ vertices, the search space therefore contains 
\[N_{base}=2^n\]
basis states.

Let $M_k$ denote the number of minimum vertex covers of cardinality $k$. The number of iterations is
\[R_{base}=R(N,M)\approx \frac{\pi}{4}\sqrt{\frac{2^n}{M_k}}.\]

\begin{table*}[t]
\centering
\caption{Practical guidelines for selecting Grover-based MVC formulations.}
\label{tab:selection_guideline}
\begin{tabular}{p{1.35\columnwidth} p{0.40\columnwidth}}
\hline
\textbf{Graph / Resource Condition} &
\textbf{Recommended Formulation} \\
\hline

Tight qubit budget &
Edge-Counting \\

Sparse or moderately sparse graph with a pronounced dense-core/low-degree-periphery structure &
Edge-Centric \\

Dense graph, large $m$, or no pronounced core-periphery structure &
Dicke-Parallel \\

\hline
\end{tabular}
\end{table*}

\paragraph{Dicke-Parallel and Edge-Counting Grover Search for MVC}
The Dicke-state formulation restricts the search to vertex subsets containing exactly $k$ selected vertices. Its search space therefore contains 
\[
N_{Dicke} = \binom{n}{k}
\]
basis states.
Since $M_k$ is the number of MVC, the number of iterations is
\[R_{dicke}=R(N,M)\approx \frac{\pi}{4}\sqrt{\frac{\binom{n}{k}}{M_k}}.\]

The Edge-Counting Grover uses the same fixed-cardinality Dicke-state search space. Therefore, 
\[R_{ECount}=R(N,M)\approx \frac{\pi}{4}\sqrt{\frac{\binom{n}{k}}{M_k}}.\]

Since
\[\binom{n}{k}\leq 2^n\]
the iteration counts satisfy
\[R_{Dicke}=R_{ECount}\leq R_{base},\]

\paragraph{Edge-Centric Grover Search for MVC}
The edge-centric formulation changes the search representation itself. Rather than representing a candidate solution by a subset of vertices, each edge $e=(u,v)$ is associated with a binary variable indicating which endpoint is selected. For a graph with $m$ edges, the edge-assignment space therefore contains
\[N_{ECentric}=2^m\]
basis states.

However, different edge assignments can induce the same minimum vertex cover. Consequently, the number of marked states can be substantially larger than the number of distinct minimum covers. As described in Section~\ref{multiplicity}, the total number of marked edge-assignment states is
\[M_{edge} = \sum_{C\in\mathcal{C}^\star} 2^{|E_{in}(C)|},\]
where $E_{in}(C)$ is the set of internal edges associated with minimum vertex cover $C$.

The direct criterion for Edge-Centric search to require fewer Grover
iterations than Dicke search is
\[
\frac{
\sum_{C \in \mathcal{C}^{\star}}
2^{|E_{{in}}(C)|}
}{
2^m
}
>
\frac{M_k}{\binom{n}{k}}.
\]

This criterion also characterizes the graph structures for which the Edge-Centric representation is most favorable. In particular, Edge-Centric search benefits when minimum vertex covers contain relatively many internal edges, since each internal edge doubles the number of edge assignments representing the same vertex cover. At the same time, the total number of graph edges should remain moderate, because the Edge-Centric search space grows as $2^m$.

Therefore, favorable instances tend to be graphs that are globally sparse or moderately sparse, while their minimum vertex covers induce relatively edge-rich subgraphs. In such cases, the representation multiplicity created by internal edges can outweigh the larger edge-based search space, producing a higher marked-state fraction than the Dicke representation and thus fewer Grover iterations. For example, a connected graph consisting of a sparse backbone with local dense structures can exhibit this behavior: the overall edge count remains moderate, while several edges lie entirely within a minimum vertex cover. Such internal edges create multiple valid edge encodings for the same cover, increasing the marked-state fraction of the Edge-Centric search.

\subsubsection{Combined Resource Considerations}

As analyzed in Section~\ref{Tdepth}, the overall logical circuit depth and
gate count scale approximately as
\[
D_{\mathrm{search}} \approx R D_{\mathrm{iter}},
\quad
G_{\mathrm{search}} \approx R G_{\mathrm{iter}},
\]
where $D_{\mathrm{iter}}$ and $G_{\mathrm{iter}}$ denote the depth and gate count of one Grover iteration, respectively, and $R$ denotes the required number of Grover iterations.

The overall resource efficiency of a formulation therefore depends jointly on its per-iteration circuit cost and its Grover iteration count. Edge-Centric is a promising choice for graphs that are globally sparse or moderately sparse but exhibit a pronounced core-periphery structure: a relatively small, densely interconnected core connected to a low-degree or pendant periphery. Such structures tend to keep the total edge count moderate while creating many alternative edge assignments around the dense core.

As the graph becomes denser and $m$ increases, however, the $2^m$ edge-assignment search space grows rapidly, reducing the iteration-count advantage of Edge-Centric. At the same time, Dicke-Parallel becomes more competitive in per-iteration depth and gate count, as observed in the dense benchmark results. Table~\ref{tab:selection_guideline} summarizes practical guidelines for selecting among the proposed Grover-based MVC formulations according to graph structure and resource constraints.
\section{Conclusion}
\label{conclusion}

In this work, we proposed resource-aware Grover-based formulations for the minimum vertex cover (MVC) problem. By exploring different choices in state-space encoding and solution verification, we provide systematic comparison of quantum MVC designs with distinct resource trade-offs.

Specifically, (i) the \textbf{Dicke-Parallel} formulation restricts the search space directly to vertex subsets of cardinality $k$, eliminating the need for explicit cardinality verification while enabling parallel edge-coverage checking; (ii) the \textbf{Edge-Counting} formulation reduces auxiliary-qubit requirements by verifying edge coverage through reversible counting, trading additional circuit depth for improved qubit efficiency; and (iii) the \textbf{Edge-Centric} formulation changes the search representation from vertex selections to edge assignments, avoiding Dicke-state preparation and enabling parallel derivation of vertex states.
This representation can substantially reduce circuit depth for sparse instances and, for graphs with a moderate total edge count and minimum vertex covers containing relatively many internal edges, can also increase the marked-state fraction through multiple edge assignments representing the same minimum vertex cover.


\bibliographystyle{IEEEtranS}
\bibliography{refs}

\end{document}